\documentclass[ceqn,10pt]{wlscirep}
\usepackage[utf8]{inputenc}
\usepackage[T1]{fontenc}
\usepackage[linesnumbered,ruled,vlined]{algorithm2e}
\usepackage{subcaption}
\usepackage{graphicx}
\usepackage{bm}
\SetKwInput{KwInput}{Input}                
\SetKwInput{KwOutput}{Output}              
\SetKwComment{Comment}{$\triangleright$\ }{}

\title{Efficient Rare Event Sampling with Unsupervised Normalising Flows}

\author[1,2,3]{Solomon Asghar}
\author[2]{Qing-Xiang Pei}
\author[1*]{Giorgio Volpe}
\author[3*]{Ran Ni}
\affil[1]{Department of Chemistry, University College London, 20 Gordon Street, WC1H 0AJ London, United Kingdom}
\affil[2]{A*STAR, Institute of High Performance Computing, Singapore, 138632, Singapore}
\affil[3]{School of Chemistry, Chemical Engineering and Biotechnology, Nanyang Technological University, Singapore, 639798, Singapore}

\affil[*]{g.volpe@ucl.ac.uk and r.ni@ntu.edu.sg}

\begin{abstract}
From Physics and Biology to Seismology and Economics, the behaviour of countless systems is determined by impactful yet unlikely transitions between metastable states known as \emph{rare events}, the study of which is essential for understanding and controlling these systems' properties. 
Classical computational methods to sample rare events remain prohibitively inefficient, and are a bottleneck for enhanced samplers requiring prior data. 
Here, we introduce a novel framework, FlowRES that uses unsupervised normalising flow neural networks to enhance Monte Carlo sampling of rare events by generating high-quality nonlocal Monte Carlo proposals.
We validate FlowRES by sampling the transition path ensembles of equilibrium and non-equilibrium systems of Brownian particles exploring increasingly complex potential surfaces. Beyond eliminating requirements for prior data, FlowRES features key advantages over established samplers: no collective variables need defining, its efficiency remains constant even as events become increasingly unlikely, and it can handle systems with multiple routes between states.
\end{abstract}

\begin{document}

\flushbottom
\maketitle

\section*{Introduction}
Rare events are low-frequency occurrences that often have significant impact on the systems manifesting them. 
Contrary to what their name suggests, they are a ubiquitous class of phenomena found in a variety of disparate systems.
Examples range from protein folding \cite{FFS_Rev7-Proteins, FFS_Rev8-Proteins2} and DNA hybridization,\cite{FFS_Rev8-DNA} to earthquakes\cite{FFS_Rev2-Earthquakes} and stock market crashes.\cite{FFS_Rev4-Stocks}
Within statistical mechanics, the most important rare events are transitions between metastable states separated by significant energetic or entropic barriers.\cite{FFS_Rev, FFS2} 
Each transition may be represented as a path -- a sequence of configurations at successive time steps. 
A correctly weighted ensemble of paths connecting the metastable states (the Transition Path Ensemble) is required to understand the transition process.\cite{TS21-ThrowingRopes}
Sampling a system’s Transition Path Ensemble is therefore critical to predicting and controlling its properties and behaviours. \cite{FFS_Rev}

Despite rapidly growing computational resources, the long waiting times that can elapse between rare events make simulation via methods based on the direct integration of dynamics equations (e.g., molecular or Langevin dynamics) highly inefficient. 
A variety of specialized techniques have thus been developed for the enhanced sampling of rare events.\cite{FFS_Rev, FFS2} Often, these techniques impose a bias potential on the system being studied,\cite{FFS_Rev16-Umbrella,FFS_Rev17-WangLandau,FFS_Rev19-Metadynamics,FFS_Rev18-AdaptiveBiasingForce,FFS2_9-TempAccel,FFS2_6-AFED} which can alter its inherent dynamics and make the generated Transition Path Ensembles inaccurate.\cite{FFS_Rev} 
Additionally, this bias must be applied along well-designed collective variables, defining which can be far from trivial.
Finding collective variables proves particularly challenging for systems where multiple routes exist between metastable states,\cite{FFS2, EfficientMultipleChannels} such as proteins which reach the same folded structure via multiple independent pathways.\cite{ProtienPaths} 
Finally, bias-potential-based samplers generally rely on microscopic reversibility and assume prior knowledge of the distribution function weighting the system's configuration space (typically, the Boltzmann distribution), limiting their applicability to only equilibrium systems\cite{EnhancedNoneq}, apart from rare exceptions.\cite{NonEqUmbrella} 

To generate accurate Transition Path Ensembles, methods that do not bias a system’s dynamics must be applied, with  the most popular (e.g.,  Transition Path Sampling -- TPS; Transition Interface Sampling -- TIS; Forward Flux Sampling -- FFS) all being Markov Chain Monte Carlo (MCMC) methods.\cite{TS21-ThrowingRopes, TS26-FFS, FFS2}
While these methods offer enhanced sampling compared to techniques based on direct integration, they lose efficiency as the events sampled become increasingly rare. 
For TPS, newly proposed paths are decreasingly likely to successfully bridge the barrier and connect the metastable states.\cite{TS21-ThrowingRopes} 
Interface based methods like TIS and FFS circumvent this issue by creating more interfaces, but this incurs a higher computational cost.\cite{TS26-FFS}
TPS additionally requires an initial path, ideally obtained by waiting for a spontaneous transition in a brute-force direct integration simulation, which can be impractical for systems where these are particularly rare.\cite{TPS_FFS_Bio} Like bias-potential-based methods, TIS and FFS also require the definition of an order parameter along some collective variables,\cite{TS26-FFS, TPSasMCMC, TPSasMCMC38-SwitchingBiochemicalNetworks} and thus struggle with systems possessing multiple routes between states.\cite{EfficientMultipleChannels}
TPS struggles with systems with multiple routes too, suffering from path trapping in the vicinity of the originally proposed sample.\cite{MetadynamicsPaths, MD16-AvoidingTraps}
This shortcoming extends to attempts made at generalising TPS to non-equilibrium systems, such as its application to active particles.\cite{TargetSearchActiveAgents}.
Of the three methods, only FFS can be straightforwardly applied to non-equilibrium systems without microscopically reversible dynamics,\cite{FFS_Rev24-TPSandRateConstants, FFS_Rev31-OnRateConstants} assuming well-designed collective variables have been found. 

Enhancing sampling using neural networks is an emerging research area within statistical physics.\cite{AdaptiveMonteCarlNormalizingFlows, AS6-MotileActiveMatterRoadmap}
Normalising flow neural networks dramatically accelerate convergence in MCMC (a ubiquitous approach to sampling) by introducing nonlocal transitions.\cite{BoltzmannGenerators, AdaptiveMonteCarlNormalizingFlows} 
However, existing frameworks necessitate standard simulations either for training networks with pregenerated samples \cite{BoltzmannGenerators} or because normalising-flow-generated data only complements standard sampling procedures.\cite{CondNormFlowsRES, AdaptiveMonteCarlNormalizingFlows}
Overall, the application of normalising flows to transition path sampling is still largely unexplored, with the few existing attempts still requiring bias potentials and the definition of collective variables.\cite{CondNormFlowsRES}

Here, we propose an unsupervised normalising flow enhanced MCMC framework for efficient rare event sampling (FlowRES -- normalising Flow enhanced Rare Event Sampler).
Unlike other approaches, FlowRES does not require standard sampling at any stage, as our network training is fully unsupervised.
Our physics-informed machine learning framework supports both equilibrium and non-equilibrium dynamics, does not require any collective variables to be defined, and is able to accurately handle systems possessing multiple routes between metastable states.
We apply our framework to several benchmark cases composed of passive and active Brownian particles exploring increasingly complex potentials. We assess FlowRES's computational efficiency and scaling by comparison with simulations via direct integration of Langevin dynamics equations, and find that, unlike other existing strategies, its efficiency does not decrease as transitions become increasingly rare.

\section*{Results}

\begin{figure*}[ht]
\centering
\includegraphics[width=\linewidth]{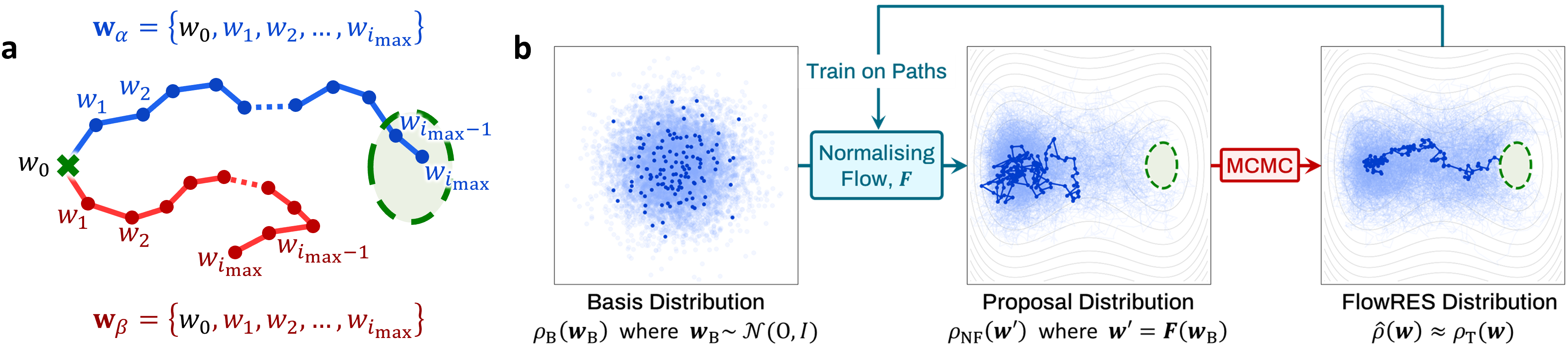}
\caption{\textbf{Transition path sampling and FlowRES workflow.} 
(\textbf{a}) Two particle's paths start from the same point $w_0$ (cross) and attempt to transition to a target region (dashed ellipse).
Each path $\textbf{w}$ is a sequence of microstates $w=(\textbf{r},\theta)$, where $\textbf{r}$ and $\theta$ are the particle's position and orientation, respectively.
Path $\textbf{w}_\alpha$ (blue) transitions successfully as its final microstate $w_{i_{\rm max}}$ is inside the target region.
Path $\textbf{w}_\beta$ (red) is non-target-reaching.
(\textbf{b}) FlowRES workflow schematic, shown for the simpler case of passive Brownian particles crossing the barrier of a double-well potential (contour plot) to reach a target region (dashed ellipse). Left: A base matrix, $\mathbf{w}_{\text{B}}$, (dark blue points) is sampled from an uncorrelated multivariate Gaussian distribution, $\rho_\text{B}(\mathbf{w}_{\text{B}})$ with $\mathbf{w}_{\text{B}} \sim \mathcal{N}(0,I)$. An ensemble of other matrices is also shown as light blue points. Centre: A normalising flow, $\textbf{\textit{F}}$, transforms $\mathbf{w}_{\text{B}}$ into a proposal path $\mathbf{w}'$ (dark blue line) shown on a background of other generated proposal paths (light blue lines), which will be used as Metropolis-Hasting proposals; their distribution, $\rho_{\text{NF}}(\mathbf{w}')$, is the proposal distribution.
Right: The proposed paths undergo a Metropolis-Hastings Monte Carlo Markov Chain (MCMC) accept-reject step (acceptance probability calculated via Eq. \ref{eq: P_acc}) to generate the FlowRES distribution $\hat{\rho}(\mathbf{w})$, which converges to the target distribution $\rho_\text{T}(\mathbf{w})$ (probability density function given by Eq. \ref{eq: PathProbEq}) with each iteration. In the schematic, the proposal is rejected as it does not reach the target region. The FlowRES distribution $\hat{\rho}(\mathbf{w})$ further trains the normalising flow and so the proposal distribution better matches the target each successive iteration.}
\label{fig:Fig 1}
\end{figure*}

\subsection*{Active Brownian Particle Model}
As a model system, we consider active Brownian particles moving with self-propulsion velocity $v$ on different conservative two-dimensional potential energy surfaces $U(\textbf{r}_{i})$,\cite{ActiveBrownianParticles} which can be described by the following discrete Langevin equations:

\begin{align}
    \textbf{r}_{i+1} \label{eq:ABP_1}
        & = \textbf{r}_{i} + v \hat{\textbf{u}}_i \Delta t - \mu \nabla U(\textbf{r}_{i})\Delta t + \sqrt{2D\Delta t} \boldsymbol{\xi}_{r}\\
    \theta_{i+1}    \label{eq:ABP_2}
        & = \theta_{i} + \sqrt{2D_{\theta}\Delta t} \,\xi_{\theta}
\end{align}

\noindent where a particle's position $\textbf{r}_{i}=(x_i,y_i)$ and orientation $\theta_i$ at time $i\Delta t$ ($\Delta t$ being the integration step) form a microstate, $w_i=(\textbf{r}_i,\theta_i$), with paths $\textbf{w} = \{w_0, w_1, ..., w_{i_{\text{max}}}\}$ being composed of $i_\textup{max}$ sequential microstates (Fig. \ref{fig:Fig 1}a). The unitary vector $\hat{\textbf{u}}_i=(\cos{\theta_i}, \sin{\theta_i})$ defines the particle's direction of motion.
$D$, $D_\theta$ and $\mu$ are its effective  translational and rotational diffusion coefficients and mobility, respectively. Translational and rotational diffusion are implemented by the noise terms $\boldsymbol{\xi}_{r}=({\xi}_{x}, {\xi}_{y})$ and $\xi_{\theta}$ (Methods). We measure energy in terms of an effective thermal energy $k_{\mathrm{B}}T_{\mathrm{eff}}:= D/\mu$ and  quantify activity with the Pe\'clet number, $\textup{Pe} = v\sqrt{3/4DD_{\theta}}$.\cite{TargetSearchActiveAgents, TS2-ActiveParticlesComplexCrowdedEnvironments} The rare events we study are transitions from a start point to a target region on various potentials (Fig. \ref{fig:Fig 1}): a standard double-well potential (Figs. \ref{fig:Fig 2} and \ref{fig:Fig 3}a-b), a double-well potential with an insurmountable wall (Fig. \ref{fig:Fig 3}c-g) and a dual-channel double-well potential (Fig. \ref{fig:Fig 4}). The values of the initial microstate $w_0=(\textbf{r}_0,\theta_0$) and of the simulation parameters (in dimensionless units) are provided in the figure captions and Methods for the various systems considered. All paths start from fixed points with uniform random orientations ($\theta_0 \sim \text{U}[-\pi,\pi]$) so the probability density function of a path, $\rho_\text{T}$, is simply the product of the probabilities of each transition between microstates (Methods), obtained from Eqs. \ref{eq:ABP_1}--\ref{eq:ABP_2} as:\cite{TargetSearchActiveAgents}

\begin{equation}\label{eq: PathProbEq}
    \rho_\text{T}(\textbf{w}) \propto \prod^{i_{\text{max}}-1}_{i=0}
    \exp\Big[{ - \frac{\big(\textbf{r}_{i+1}-\textbf{r}_i-v\hat{\textbf{u}}_i\Delta t +\mu\nabla U(\textbf{r}_i)\Delta t\big)^2}{4D\Delta t}}\Big]
    \exp\Big[{\cos\Big(\frac{\theta_{i+1} - \theta_{i}}{\sqrt{2D_{\theta}\Delta t}}\Big)}\Big]
\end{equation}

\noindent This probability density function defines the target distribution we sample from: calculating the value of this function for any given path $\textbf{w}$ yields its probability in the distribution of transition paths, which we use to train FlowRES neural network to generate correctly weighted transition paths in an unsupervised manner. For passive particles we set $v=0$ and remove the second exponential term from Eq. \ref{eq: PathProbEq}.

\subsection*{FlowRES}
FlowRES efficiently samples rare events from a system’s transition path ensemble using Metropolis-Hastings MCMC with each new proposal generated by a normalising flow neural network (Methods, Fig. \ref{fig:Fig 1}b).
Our network is implemented with affine coupling transformations\cite{RealNVP} that use WaveNet\cite{WaveNet} (a network for efficiently processing time series) as part of their architecture (Methods, Figs. \ref{fig:SupShematic_PosOnly} - \ref{fig:SupShematic_Full}). 

FlowRES samples using $c_\text{total}$ parallel Markov chains $\{{\textbf{w}_c(m)}\}$, where $c \in \{1, 2,..., c_\text{total}\}$ identifies each chain and $m \in \{0, 1,..., m_\text{max}\}$ is the sampling iteration. 
Each chain’s initial path, $\textbf{w}_c(0)$, is a simple random walk, i.e. a passive Brownian trajectory with randomly evolving orientation in the absence of any external potential (Eqs. \ref{eq:ABP_1}--\ref{eq:ABP_2} with $v=0$ and $U(\textbf{r}_{i})=0$).
These dynamics are very cheap to simulate and resemble the target dynamics, accelerating network training convergence. At each successive sampling iteration $m$, our network proposes $c_\text{total}$ new paths so that each chain may grow by one path.
To generate a proposal path, the normalising flow network takes as input a matrix, $\textbf{w}_\text{B}$, of the same shape as the path we aim to generate, $i_\text{max} \times n$.
For passive particles ($v = 0$, $n = 2$), $\textbf{w}_\text{B}$ is an $i_\text{max} \times 2$-matrix of random values sampled from an uncorrelated multivariate Gaussian distribution. 
For active particles ($v \neq 0$, $n = 3$), $\textbf{w}_\text{B}$ is obtained by horizontally concatenating such a Gaussian matrix with an $i_\text{max} \times 1$-sequence of independently pre-generated orientations following Eq. \ref{eq:ABP_2}. Our normalising flow, $\textbf{\textit{F}}$, takes $\textbf{w}_\text{B}$ as input and outputs a proposal path $\textbf{w}'=\textbf{\textit{F}}(\textbf{w}_\text{B})$ (Fig. \ref{fig:Fig 1}b, Algorithm \ref{algo: Alg 1}). 
For passive particles, $\textbf{\textit{F}}$ transforms all of $\textbf{w}_\text{B}$ (Fig. \ref{fig:SupShematic_PosOnly}).
For active particles, the flow uses the orientation components of $\textbf{w}_\text{B}$ to condition the transformation of its position components, but leaves the orientations themselves unchanged (Fig. \ref{fig:SupShematic_Full}). 

For each chain $c$, a new proposal path $\textbf{w}_c'= \textbf{\textit{F}}(\textbf{w}_{\text{B},c})$ is added to the chain following a Metropolis-Hastings MCMC accept-reject step \cite{TPSasMCMC} with an acceptance probability:

\begin{align}\label{eq: P_acc}
P_\textup{acc}(\textbf{w}_c(m),\textbf{w}'_c) = \begin{cases}            0    & \text{{if $\textbf{w}_c(m)$ is target-reaching but $\textbf{w}_c'$ is not}}  \\
                                         1 & \text{if $\textbf{w}_c(m)$ is not target-reaching but $\textbf{w}_c'$ is } \\
                              \min\Big[1, \frac{\rho_\textup{T}(\textbf{w}_c')\rho_\textup{NF}(\textbf{w}_c(m))}
                                           {\rho_\textup{T}(\textbf{w}_c(m))\rho_\textup{NF}(\textbf{w}_c')}\Big]
                                           & \text{if $\textbf{w}_c(m)$ and $\textbf{w}_c'$ are both target-reaching or neither is target-reaching}\end{cases}
\end{align}

\noindent where $\rho_\text{NF}(\textbf{w})$ is the probability for the normalising flow at the current iteration to generate a given path $\textbf{w}$ (Methods, Fig. \ref{fig:Fig 1}b). As we aim to generate only transition paths, our acceptance probability $P_\textup{acc}(\textbf{w}_c(m), \textbf{w}'_c)$ depends on whether the current path in the Markov chain, $\textbf{w}_c(m)$, and the proposed next path, $\textbf{w}'_c$, reach the target region (Fig. \ref{fig:Fig 1}b).
If $\textbf{w}_c(m)$ is target-reaching but $\textbf{w}'_c$ is not, then $P_\text{acc}=0$ and the proposal is instantly rejected. If $\textbf{w}_c(m)$ is not target-reaching but $\textbf{w}'_c$ is, then $P_\text{acc}=1$ and $\textbf{w}'_c$ is accepted.
These two conditions ensure that the number of target-reaching paths increases monotonically. In all other cases, we calculate $P_\text{acc}$ based on the probabilities of finding each path in the target distribution, $\rho_\text{T}$, and of the normalising flow generating each path, $\rho_\text{NF}$ (Methods), and $\textbf{w}'_c$ is accepted with a probability $P_\text{acc}$. If the proposal is accepted, we set ${\textbf{w}_c(m+1)}= \textbf{w}'_c$; alternatively, ${\textbf{w}_c(m+1)}= \textbf{w}_c(m)$ (Algorithm \ref{algo: Alg 1}). After this accept-reject step, we have a new ensemble $\{\textbf{w}_c(m+1)\}_{c=1}^{c_\textup{total}}$ to train our normalising flow on, updating the network weights $\psi$ with a maximum-likelihood loss function $\mathcal{L}(\psi)$ (Methods). Note that unlike other transition path sampling schemes,\cite{TargetSearchActiveAgents, TPSasMCMC} our accept-reject step does not always reject non-target-reaching paths.
In early sampling iterations, a large share of chains may be composed of mostly non-target-reaching paths, yet, because of our $P_\textup{acc}$, even paths from these chains will get more physically realistic at each iteration; while these paths may not contribute to training the network to generate successful target-reaching paths, they still help it learn how to generate more realistic paths. At each iteration, $\{\textbf{w}_c(m+1)\}_{c=1}^{c_\textup{total}}$ more closely matches the target distribution, with paths becoming more physically realistic and a larger portion of paths being target-reaching.
As the normalising flow trains on $\{\textbf{w}_c(m+1)\}_{c=1}^{c_\textup{total}}$, each iteration's proposals are better than the previous, and the FlowRES-generated distribution converges to the target distribution ($\hat{\rho}(\textbf{w}) \approx \rho_{\rm T}(\textbf{w})$) as $m$ grows, with the $m_\text{max}$-th distribution closely matching the target. 

\begin{algorithm}[!htb]
\DontPrintSemicolon
    \KwInput{$\textbf{\textit{F}}$ initial network,
             $\{\textbf{w}_c(0)\}_{c=1}^{c_\textup{total}}$ initial chains,
             $P_\textup{acc}$ acceptance probability function (Eq. \ref{eq: P_acc}),
             $m_{\textup{max}}$ number of iterations}
    \KwOutput{$\textbf{\textit{F}}$ trained network,
              $\{\textbf{w}_c(m_\textup{max})\}_{c=1}^{c_\textup{total}}$ equilibrated chains}
    $m = 0$ 
    
    \While(\Comment*[f]{Until $m_\textup{max}$ iterations pass}){$m < m_{\textup{max}}$} 
    { 
        \For(\Comment*[f]{For each chain $c$}){$c=1,...,\ c_\textup{total}$ }
        {
            $\textbf{w}_{\text{B},c} \sim \rho_{\text{B}}$ \Comment*[f]{Sample a base matrix}
            
            $\textbf{w}'_c = \textbf{\textit{F}}(\textbf{w}_{\text{B},c})$ \Comment*[f]{Propose a path}

            $u \sim \text{U}[0,1]$ \Comment*[f]{Generate uniform random number}
            
            \uIf(\Comment*[f]{Accept or reject proposed path}){$P_\textup{acc}(\textbf{w}_c(m), \textbf{w}'_c) > u$} 
                {
                $\textbf{w}_c(m+1) = \textbf{w}'_c$
                }
            \Else
                {
                $\textbf{w}_c(m+1) = \textbf{w}_c(m)$ 
                }
        }
    
        Train $\textbf{\textit{F}}$ with loss $\mathcal{L}(\psi)$ on samples $\{\textbf{w}_c(m+1)\}_{c=1}^{c_\textup{total}}$ \Comment*[f]{Train on new path ensemble}
        
        $m = m + 1$
    }
\caption{FlowRES - normalising Flow enhanced Rare Event Sampler}\label{algo: Alg 1}
\end{algorithm}

\subsection*{Enhanced Sampling of Passive Systems}

\begin{figure*}[ht!]
\centering
\includegraphics[width=0.9\linewidth]{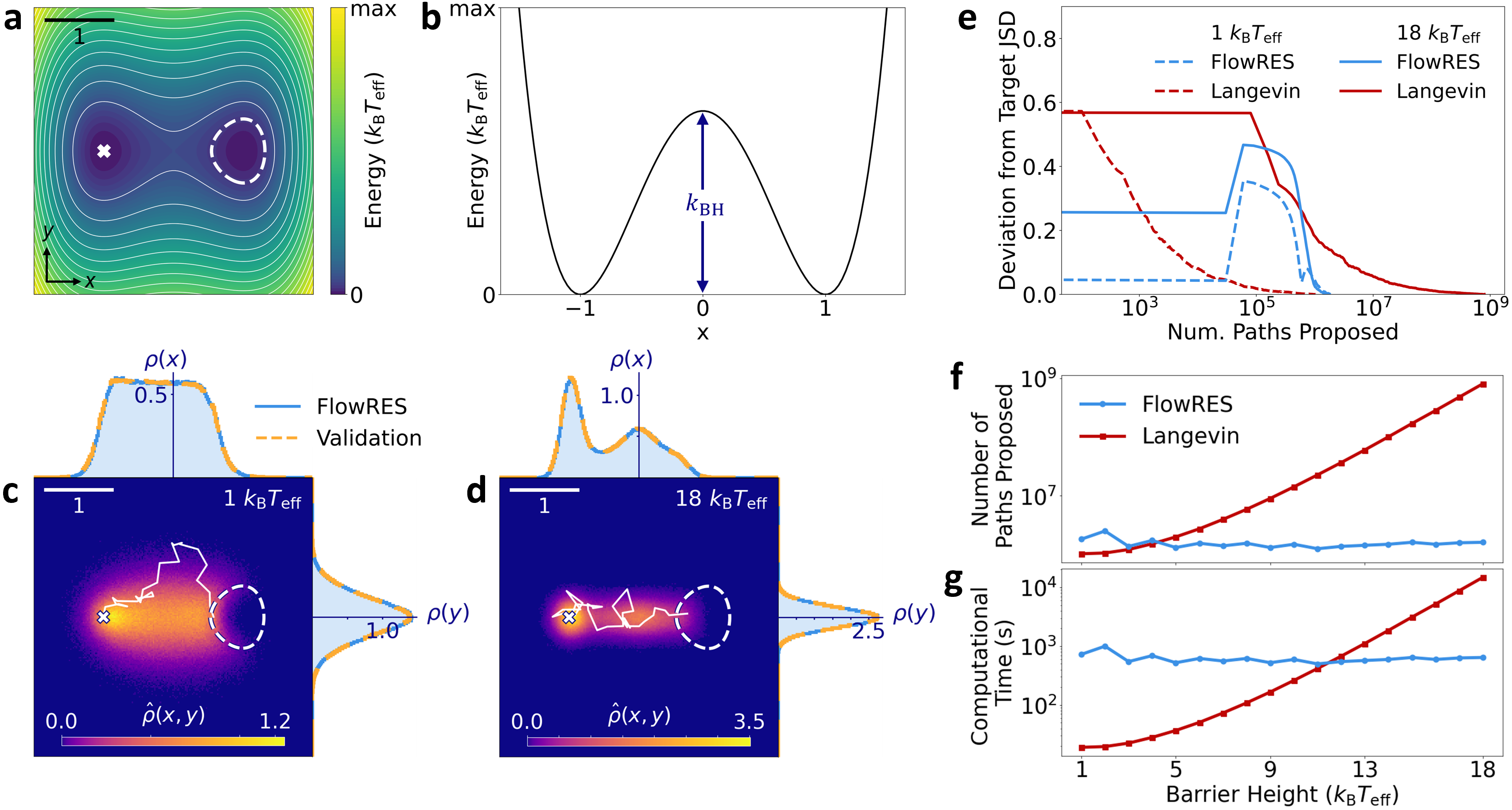}
\caption{\textbf{Barrier crossing of a Brownian particle in a double-well potential.}
({\bf a}) A double-well potential (Methods). The particle’s starting point (white cross) is separated from the target region (enclosed by dashed line) by an energy barrier.
Unitary scale bar: upper left; reference frame: lower left.
({\bf b}) Potential profile along $y=0$ with barrier height highlighted (arrow). The parameter $k_{\rm BH}$ controls the height of the barrier (Methods).
({\bf c-d}) FlowRES-generated probability densities for target-reaching paths $\hat{\rho}(x,y)$ (colour map) for barrier heights of (\textbf{c}) 1 $k\textsubscript{B}T\textsubscript{eff}$ and (\textbf{b}) 18 $k\textsubscript{B}T\textsubscript{eff}$, flanked by FlowRES-generated marginal probability densities $\hat{\rho}(x)$ and $\hat{\rho}(y)$ (shaded blue histograms) and validation marginal probability densities $\rho_{\rm V}(x)$ and $\rho_{\rm V}(y)$ (dashed orange histograms) generated via direct integration of Eq. \ref{eq:ABP_1}; both distributions correspond closely. Example paths are shown in white. Simulation parameters:
$\Delta t=0.05$, $\mu=0.1$, $D=0.15$ and $i_\textup{max}=32$ (Methods).
({\bf e}) Deviations of the Jensen Shannon distance (JSD) between validation ensemble and ensembles of target-reaching paths generated with FlowRES (blue lines) and direct integration (red lines) from the target JSD (Methods) as a function of number of proposed paths for two barrier heights: $1 \ k\textsubscript{B}T\textsubscript{eff}$ (dashed lines) and $18 \ k\textsubscript{B}T\textsubscript{eff}$ (solid lines). ({\bf f}) Total number of paths proposed to yield ensembles with zero deviation from target JSD via FlowRES (blue line) and direct integration (red line) and ({\bf g}) corresponding computational time
as a function of barrier height.
The computational cost of direct integration of Langevin equations increases exponentially with barrier height, while FlowRES maintains near constant computational cost. 
}
\label{fig:Fig 2}
\end{figure*}

We begin by using FlowRES to explore a two-dimensional system of passive Brownian particles exploring a double-well potential with constant topology but variable barrier height (Figs. \ref{fig:Fig 2}a-b, Methods).
We set $c_\text{total}=30,000$ and $m_\text{max}=100$, selecting a $c_\text{total}$ that provides efficient sampling and an $m_\text{max}$ that comfortably allows for convergence to the target distribution.
Particles start at the minimum of the left well and have a small probability of transitioning to the right well. The spatial path probability density plots in Figs. \ref{fig:Fig 2}c-d at two barrier heights ($1 \ k\textsubscript{B}T\textsubscript{eff}$ and $18 \ k\textsubscript{B}T\textsubscript{eff}$) show that FlowRES generates realistic paths that capture the same statistics as standard simulations based on direct integration of Eq. \ref{eq:ABP_1}.

While the efficiency of direct integration can be evaluated by measuring the computational cost of generating $N$ target-reaching paths, we cannot directly apply such a metric to Monte Carlo methods.
Within its first few iterations, FlowRES generates numerous target-reaching, yet physically unrealistic paths, and gradually improves physical realism at each iteration. 
Assessing FlowRES' efficiency requires a metric that considers the quality of the generated ensemble rather than the sheer quantity of its paths. To do so, we use the Jensen-Shannon distance (JSD) from a validation ensemble of 50,000 target-reaching paths generated via direct integration of Eq. \ref{eq:ABP_1} (Methods). Specifically, at each iteration we calculate the deviation between this JSD for the FlowRES-generated ensemble and a target JSD between 10,000 target-reaching direct integration paths and the validation ensemble (Methods). After this deviation reaches zero, we can conclude that the FlowRES-generated ensemble is at least as statistically accurate as that obtained by direct integration and may compare the computational costs of the two methods. Fig. \ref{fig:Fig 2}e shows how this JSD deviation evolves for FlowRES and, as a point of comparison, for direct integration with the number of paths proposed for barrier heights of $1 \ k_{\mathrm{B}}T_{\mathrm{eff}}$ and $18 \ k_{\mathrm{B}}T_{\mathrm{eff}}$. For the system with the lower barrier height, both methods require a similar number of proposals (direct integration: $1.1\times 10^6$; FlowRES: $1.9\times 10^6$), with direct integration being slightly more efficient. 
When the barrier height is increased, transitions become much rarer and
the number of proposals required by direct integration increases by about three orders of magnitude ($8\times10^8$).
FlowRES efficiency, however, is not effected by this increase in event rarity, with the higher barrier system requiring no more proposals  ($1.7\times10^6$) than the lower barrier one. This constant efficiency is possible as learning how to directly generate transition paths is a purely statistical task, so the barrier heights of potentials have no direct bearing on its computational cost.

Fig. \ref{fig:Fig 2}f plots how the number of proposals required for the aforementioned JSD deviation to reach zero changes as a function of barrier height, for both FlowRES and direct integration. FlowRES proves more efficient than direct integration for barrier heights over $4 \ k_{\mathrm{B}}T_{\mathrm{eff}}$.
As expected, direct integration loses efficiency exponentially with increasing barrier height, while FlowRES maintains constant efficiency, always requiring $\sim 10^6$ proposals. Fig. \ref{fig:Fig 2}g shows the effects of barrier height in terms of computational time required on our computational resources (Methods). 
FlowRES outperforms direct integration from barrier heights of $10 \ k_{\mathrm{B}}T_{\mathrm{eff}}$ onward.
Although the generation time per proposal is longer for FlowRES than for direct integration, FlowRES still outperforms direct integration from barrier heights of $10 \ k_{\mathrm{B}}T_{\mathrm{eff}}$ onward. 
FlowRES's strength lies with more intelligently designed proposals which allow FlowRES to generate physically realistic transition path ensembles with a computational efficiency that, unlike other enhanced samplers (e.g., FFS, TIS, TPS),\cite{TS26-FFS,TS21-ThrowingRopes} remains constant as event rarity increases. 

\subsection*{Extending FlowRES to Active Systems}

\begin{figure*}[ht]
\centering
\includegraphics[width=0.8\linewidth]{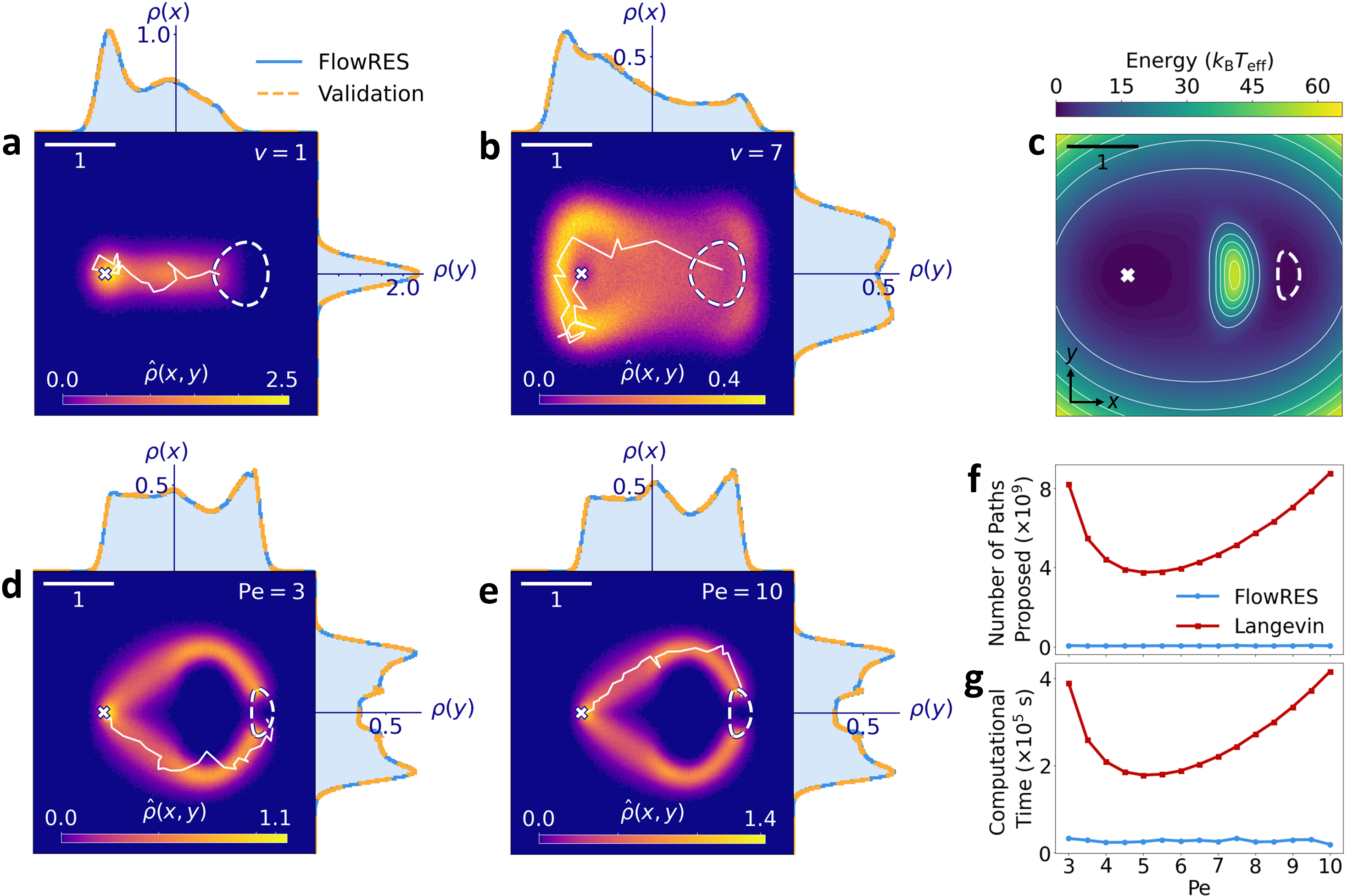}
\caption{\textbf{Enhanced sampling of non-equilibrium active Brownian systems.} 
({\bf a-b}) FlowRES-generated probability density $\hat{\rho}(x,y)$ (colour map) for target-reaching paths of active Brownian particles self-propelling at a speed of ({\bf a}) $v = 1$ and ({\bf b}) $v = 7$ in a double-well potential (Fig. \ref{fig:Fig 2}a-b, barrier height of $15 \ k_{\mathrm{B}}T_{\mathrm{eff}}$), flanked by FlowRES-generated marginal probability densities $\hat{\rho}(x)$ and $\hat{\rho}(y)$ (shaded blue histograms) and validation marginal probability densities ${\rho}_{\rm V}(x)$ and ${\rho}_{\rm V}(y)$ (dashed orange histograms) generated via direct integration. Paths start from the white cross and travel to the target region enclosed by the white dashed line. Example paths are shown in white. Simulation parameters as in Fig. \ref{fig:Fig 2} with $D_\theta=2.5$. Unitary scale bar: upper left.
({\bf c}) Double-well potential with an insurmountable wall separating the starting point (white cross) and target region (enclosed by dashed line) (Methods). Reference frame: lower left.
({\bf d-e}) FlowRES-generated probability densities and comparison with direct integration as in {\bf a-b} for active particles moving on the potential in {\bf c} with Pe\'clet number ({\bf d}) $\textup{Pe} = 3$ and ({\bf e}) $\textup{Pe} = 10$, obtained varying the rotational diffusion coefficient $D_\theta$ at fixed $v = 2$.   
All other simulation parameters as in {\bf a-b} but $\Delta t=0.025$. 
({\bf f}) Total number of paths proposed to yield ensembles with zero
deviation from target JSD (Methods) via FlowRES (blue line) and direct integration (red line) and ({\bf g}) corresponding computational time  as a function of Pe\'clet number.
The computational cost of direct integration of Langevin equations changes with Pe\'clet number, while FlowRES maintains near constant computational cost. 
}
\label{fig:Fig 3}
\end{figure*}

Introducing activity to the previous system radically changes the structure of the transition path ensemble, with paths exhibiting counter-intuitive search routes. \cite{TargetSearchActiveAgents} Nonetheless, FlowRES (with $c_\text{total}=50,000$ and $m_\text{max}=300$) still generates realistic distributions of positions (Figs. \ref{fig:Fig 3}a-b) and orientations (Figs. \ref{fig:SupABPActive}) for the transition path ensembles of active particles moving at different velocities: $v = 1$ (Figs. \ref{fig:Fig 3}a and \ref{fig:SupABPActive}a) and $v = 7$ (Figs. \ref{fig:Fig 3}b and \ref{fig:SupABPActive}b). Across this wide activity range, our method generates the same statistics as direct integration and captures the different routes that active paths may take: while low activity (Fig. \ref{fig:Fig 3}a) particles, as passive ones (Fig. \ref{fig:Fig 2}), take short energy-minimising routes, particles with higher activity are driven by their self-propulsion to climb the potential surface away from the target before traveling toward it along higher energy regions. As for passive particles (Fig. \ref{fig:Fig 2}f-g), FlowRES shows close-to constant computational cost (unlike direct integration) simulating active systems as barrier height grows (Fig. \ref{fig:SupABPActive_scale}).

A potential where the starting point and target region are separated by an insurmountable wall presents a more challenging system for active particles to traverse (Fig. \ref{fig:Fig 3}c, Methods).
Particles cannot simply cross over this obstacle as they did the barrier of the double-well potential (Fig. \ref{fig:Fig 3}a-b) and must instead reach the target by navigating around the wall (Fig. \ref{fig:Fig 3}d-e). This task is complicated by the  motility-induced tendency of active particles to localize near steep walls.\cite{TargetSearchActiveAgents, TS2-ActiveParticlesComplexCrowdedEnvironments} As persistence increases, particles tend to spend longer at the steep walls of the obstacle and of the double-well potential. This motility-induced effect makes the transition to the target region very rare, so standard simulations can be highly inefficient. By varying the rotational diffusion coefficient $D_\theta$ at fixed $v$, we alter the system's Pe\'clet number between $3$ and $10$, exploring the effects of particle's persistence on FlowRES efficiency  (with $c_\text{total}=100,000$ and $m_\text{max}=750$).
Once again, from the distributions of positions (Figs. \ref{fig:Fig 3}d-e) and
orientations (Fig. \ref{fig:SupABPBlock}) of the transition path ensembles, we see that FlowRES generates accurate paths for active particles on this more complex potential. FlowRES is more efficient than direct integration, both in terms of number of proposed paths (Fig. \ref{fig:Fig 3}f) and computational time (Fig. \ref{fig:Fig 3}g). Using direct integration, a minuscule proportion of paths are target-reaching at lower persistence, thus requiring high numbers of proposals (e.g., $8.2\times10^9$ at $\textup{Pe}=3$ ) to generate significant statistics:
many low persistence particles do not explore far enough to reach the target, instead spending time in the left well by the starting point. Simulating high persistence particles is also challenging for direct integration (e.g., $\textup{Pe}=10$ requires $8.8\times10^9$ proposals to generate 10,000 target-reaching paths), as they tend to spend long periods of time near boundaries,\cite{TargetSearchActiveAgents, TS2-ActiveParticlesComplexCrowdedEnvironments} with very few paths ever reaching the target zone. At intermediate values of persistence ($\textup{Pe}=5$), direct integration is less inefficient (requiring only $3.7\times10^9$ proposals for 10,000 target-reaching paths) as particles' dynamics strike a good balance between surface exploration and reduced localization at the boundaries. Nonetheless, in all cases, FlowRES outperforms direct integration as it requires an average of only $5.0\times10^7$ proposals to generate equivalent statistics -- at least two orders of magnitude less than the best case with direct integration. As before, the rarity of transitions (determined by the Pe\'clet number in this case) does not alter FlowRES efficiency.
This also emerges when comparing efficiency in terms of computational time (Fig. \ref{fig:Fig 3}g), as FlowRES takes approximately 7.5 hours to generate a statistically accurate transition path ensemble for this system, i.e. between 6 and 22 times faster than direct integration.

\subsection*{Systems with Multiple Routes}

\begin{figure*}[ht]
\centering
\includegraphics[width=0.85\linewidth]{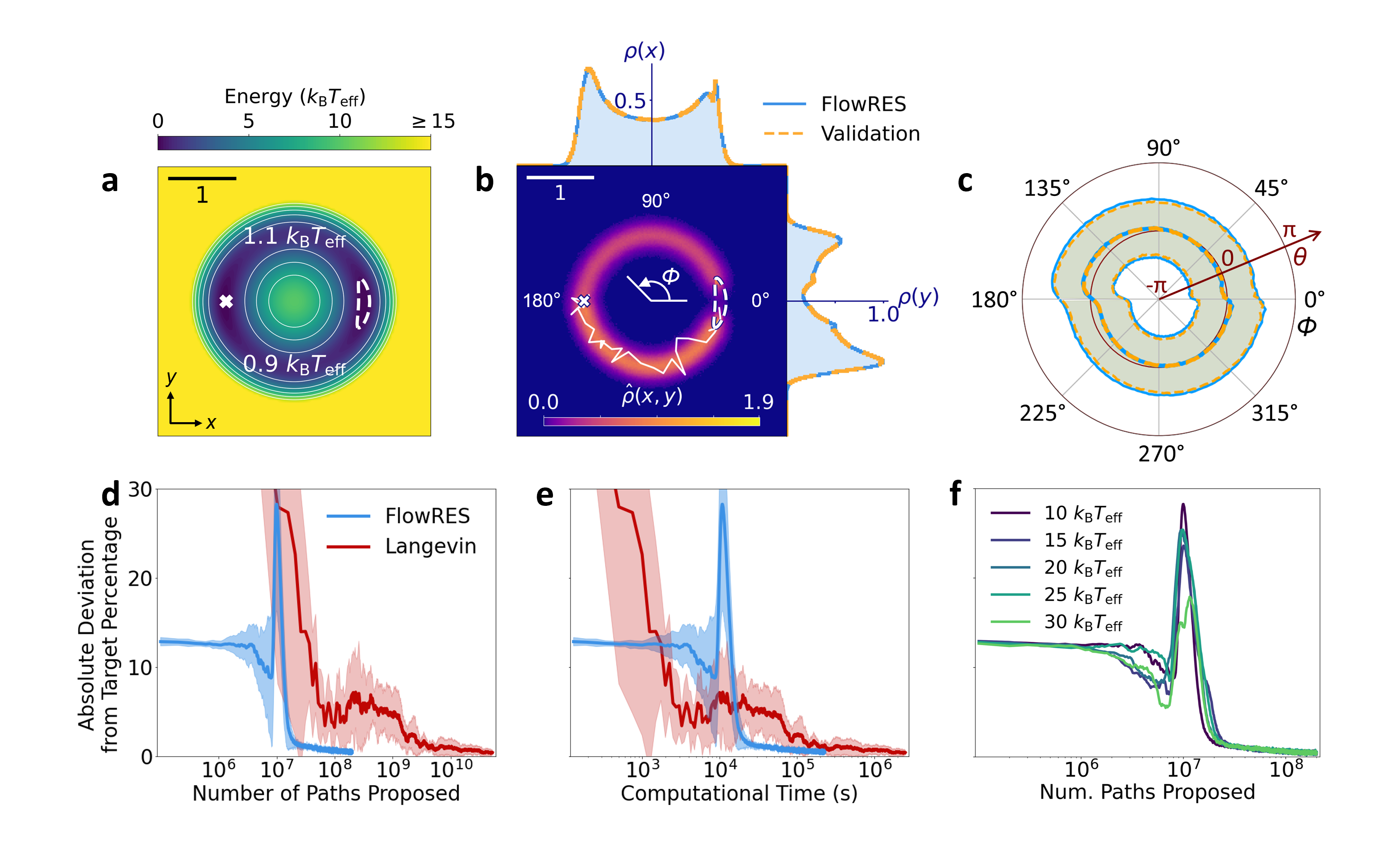}
\caption{\textbf{Finding the route ratio for an asymmetric dual-channel double-well potential.}
({\bf a}) Asymmetric dual-channel double-well potential (Methods). The start point (white cross) is connected to the target region (enclosed by dashed line) by two distinct channels with different barrier heights. Unitary scale bar: upper right; reference frame: lower left.
({\bf b}) 
FlowRES-generated probability density $\hat{\rho}(x,y)$ (colour map) for target-reaching paths of active Brownian particles on the dual-channel double-well potential in {\bf a}, flanked by FlowRES-generated marginal probability densities $\hat{\rho}(x)$ and $\hat{\rho}(y)$ (shaded blue histograms) and validation marginal probability densities ${\rho}_{\rm V}(x)$ and ${\rho}_{\rm V}(y)$ (dashed orange histograms) generated via direct integration. An example path taking the low-energy route is shown in white. Simulation parameters: $\Delta t=0.05$, $v=0.15$, $\mu=0.25$, $D=0.1$, $D_{\theta}=1$ and $i_\textup{max}=32$. The polar angle $\phi$ from the centre of the potential is labelled at $0^\circ$, $90^\circ$ and $180^\circ$. 
({\bf c}) Average orientation $\overline{\theta}$ (solid lines) and corresponding standard deviation (shaded region) for active particles at different polar angles $\phi=\arctan(\frac{x}{y})$ on the potential for FlowRES (blue) and direct integration (red) of Langevin equations (Eqs. \ref{eq:ABP_1} and \ref{eq:ABP_2}). For clarity polar angles use degrees while particle orientations use radians.
({\bf d-e}) Average deviation of calculated percentages of top channel occupancy from the target value ($36.71\%$, calculated from 50,000 direct integration paths) against (\textbf{d}) number of paths proposed and (\textbf{e}) computational time for FlowRES (blue line) and direct integration (red line). Shaded areas represent standard deviations around the average from five independent simulations. 
({\bf f}) Same as {\bf d} for FlowRES only, calculated for several heights of the peak separating the two channels in the potential in {\bf a}. In {\bf d-f}, the peak around $10^7$ proposed paths ($10^4$ s in {\bf e}) is because, at this point in training, the network generates mostly paths from one of the two channels, before further iterations allow for more mixing between the two routes.
}
\label{fig:Fig 4}
\end{figure*}

To evaluate how FlowRES performs on landscapes with multiple routes, we consider a potential with two distinct channels separating the start point from the target region: a top route with a $1.1\ k_\mathrm{B}T_\mathrm{eff}$ barrier and a bottom route with a $0.9\ k_\mathrm{B}T_\mathrm{eff}$ barrier (Fig. \ref{fig:Fig 4}a, Methods). The two routes are separated by a central peak of 10 $k_\mathrm{B}T_\mathrm{eff}$.
Due to the difference in barrier height along each route, we can expect a higher proportion of paths to reach the target via the lower-energy bottom route. 
Calculating what ratio of paths takes each route is difficult with existing enhanced samplers, with TPS-based method suffering from path trapping\cite{MetadynamicsPaths, MD15-OnEfficientPS, MD16-AvoidingTraps} and interface-based methods (TIS and FFS) requiring the definition of collective variables.\cite{FFS2, EfficientMultipleChannels}

Analysing a validation ensemble of 50,000 target-reaching paths generated via direct integration yields a target value of 36.71\% for the percentage of paths crossing via the higher-energy top channel. Like previous cases (Figs. \ref{fig:Fig 2}-\ref{fig:Fig 3}), FlowRES (with $c_\text{total}=100,000$ and $m_\text{max}=2,000$) captures the same statistics as direct integration both in terms of the probability densities of the particles' positions (Fig. \ref{fig:Fig 4}b) and orientations (Figs. \ref{fig:Fig 4}c and \ref{fig:SupABPTCDW}a). As more paths are proposed, FlowRES' initially random estimate for the percentage of paths in the higher-energy channel converges to this target value (Figs. \ref{fig:Fig 4}d and \ref{fig:SupABPTCDW}b in terms of JSD deviation) after far fewer proposals than those required by direct integration, reaching $36.71 \pm0.5 \%$ within $7.98\times10^7$ proposals while direct integration required 400 times as many proposals ($3.26\times10^{10}$) to reach the same accuracy.
In terms of computational time (Figs. \ref{fig:Fig 4}e and \ref{fig:SupABPTCDW}c in terms of JSD deviation), FlowRES reaches this target value 17 times faster than direct integration. 
In the generated transition path ensemble, as intuition would suggest, the preferred orientations for a particle to explore a channel follow the channel's curvature, as shown in Fig. \ref{fig:Fig 4}c where the orientations are plotted as a function of the polar angle coordinate $\phi=\arctan(\frac{x}{y})$: the paths exploring the high-energy channel at the top of Fig. \ref{fig:Fig 4}a start with an average positive orientation that reduces to $\bar\theta = 0$ by the time the particle is halfway to the target ($\phi=90^\circ$, $x = 0$), with negative orientations preferred thereafter until the target is reached; the lower-energy channel at the bottom of Fig. \ref{fig:Fig 4}a follows the opposite trend. Finally, for calculations relating to the ratio of occupancies of different routes, existing enhanced samplers typically decrease in efficiency as the barrier between these routes is increased as it makes mixing between them less and less likely.\cite{MetadynamicsPaths, MD15-OnEfficientPS, MD16-AvoidingTraps}
Interestingly, FlowRES does not suffer from such a limitation as the number of proposals required is independent of barrier height (Fig. \ref{fig:Fig 4}f).
 
\section*{Discussion}

We have developed a novel simulation framework, FlowRES, that uses normalising flow neural networks to enhance MCMC sampling of rare events.
We apply this framework to equilibrium and non-equilibrium systems of Brownian particles and efficiently generate transition path ensembles for both. 
As it uses fully unsupervised networks, unlike other enhanced paths samplers or normalising flow augmented MCMC schemes FlowRES requires no pregenerated target ensemble samples from prior simulation. 
In contrast to simulations via direct integration or existing enhanced schemes, FlowRES experiences no slowdown as event rarity increases. This is demonstrated in cases where events become exceedingly rare due to growing energetic barriers and due to motility-induced topological localization. Unlike other path samplers, no collective variables or reaction coordinate need to be defined. Additionally, FlowRES does not struggle with systems possessing multiple routes between metastable states, and unlike other methods,\cite{EfficientMultipleChannels} is not sensitive to increases in the height of barriers between these routes. Future development for FlowRES could expand the scope of what can be simulated by further developing the normalising flow network architecture used e.g., by employing architectures based on rational quadratic spline functions\cite{RQS} instead of affine coupling layers. Another interesting line of future inquiry would be to extend the FlowRES framework to more complex topological spaces (e.g. tori or spheres) by building new flows accordingly.\cite{FlowsToriSphere}
Additionally, FlowRES could be adapted to study active target search optimization problems and enable efficient characterisation of different behavioural policies.\cite{volpe2017topography,ABP_search_unknown_pos, Path_Plan_turbulent, Reinforcement_Complex_motility, Optimal_Foraging, Find_Targets_Micro_World}
In conclusion, FlowRES offers a new approach to path sampling, free from many of the limitations affecting other rare event sampling methods. We envision that our framework can therefore help study many currently inaccessible rare events of interest in statistical mechanics and beyond.

\section*{Methods}

\subsection*{Probability Density Function of a Path}

In Eq. \ref{eq:ABP_1}, we model translational diffusion, $\boldsymbol{\xi}_{r}=({\xi}_{x}, {\xi}_{y})$, using two white noise terms, ${\xi}_{x}$ and ${\xi}_{y}$, sampled from a Gaussian distribution with zero mean and unit variance.\cite{volpe2014simulation} 
In Eq. \ref{eq:ABP_2}, we model rotational diffusion, $\xi_{\theta}$, similarly but sample from a von Mises distribution with zero mean and unit concentration, which closely approximates a normal distribution wrapped around the unit circle but is more tractable.\cite{Bessel} 
Rearranging Eqs. \ref{eq:ABP_1} and \ref{eq:ABP_2} for $\boldsymbol{\xi}_{r}$ and $\xi_{\theta}$ respectively yields equations for recovering the noise terms from a transition $w_i \rightarrow w_{i+1}$.
Applying this to all the transitions in a path yields all the noise values associated with that path. 
The probability of all the positional noise can be calculated using a Gaussian probability density function. 
Similarly, applying a von Mises probability density function to all the orientation noise values yields their probability. 
The product of these probabilities is the probability density function of the paths, i.e., $\rho_{\rm T}(\textbf{\text{w}})$ shown in Eq. \ref{eq: PathProbEq}.

\subsection*{Potential Energy Surfaces and Simulation Parameters}
We consider three main potential energy surfaces: a standard double-well potential (Figs. \ref{fig:Fig 2} and \ref{fig:Fig 3}a-b), a double-well potential with an insurmountable wall (Fig. \ref{fig:Fig 3}) and a dual-channel double-well potential (Fig. (Fig. \ref{fig:Fig 4}). We set dimensionless simulations parameters in terms of effective energy ($k_{\mathrm{B}}T_{\mathrm{eff}}:= D/\mu$) and activity (Pe\'clet number, $\textup{Pe} = v\sqrt{3/4DD_{\theta}}$).\cite{TargetSearchActiveAgents, TS2-ActiveParticlesComplexCrowdedEnvironments}

\subsubsection*{Double-Well Potential} 
Figs. \ref{fig:Fig 2} and \ref{fig:Fig 3}a-b use a double-well potential, shown in Fig. \ref{fig:Fig 2}a:
\begin{equation}
    \centering
    U(x, y) = \frac{k_{\text{BH}}}{2}\Big(k_{x}(x^2 - x_0^2)^2 + \frac{k_y}{2} + y^2\Big)
\end{equation}
The parameters $k_{x}$ and $k_{y}$ define the steepness of the potential in the $x$ and $y$ directions, respectively. Both minima lie at $y = 0$, and their $x$ coordinates are dictated by $x_0$, with one minima at $(-x_0, 0)$ and the other at $(x_0, 0)$. We set $k_x=2$, $k_y=10$ and $x_0=1$, with $k_{\text{BH}}$ varying between $1$ and $18$. Particles start at $(-1, 0)$ and are considered to have successfully transitioned to the other basin if they reach regions of space where $x > 0$ and $U(x, y) \leq \frac{k_{\text{BH}}}{2}$ (Fig. \ref{fig:Fig 2}a). The dynamics of the particles exploring this surface are described by Eqs. \ref{eq:ABP_1} and \ref{eq:ABP_2}. We set $\Delta t=0.05$, $\mu=0.1$, $D=0.15$, and $i_\textup{max}=32$. Fig. \ref{fig:Fig 2} considers passive particles, so $v=0$ and the rotational dynamics do not influence the translational degrees of freedom and can be neglected. The systems in Fig. \ref{fig:Fig 3}a-b are active; we set $D_\theta=2.5$ with $v=1$ (Fig. \ref{fig:Fig 3}a) or $v=7$ (Fig. \ref{fig:Fig 3}b).

\subsubsection*{Double-Well Potential with a Wall}
This potential, shown in Fig. \ref{fig:Fig 3}c and used in Fig. \ref{fig:Fig 3}d-g, consists of two harmonic potential wells separated by a curved wall: 

\begin{equation} \label{eq:Walls}
    U(x, y) = U_{\textup{LW}} \times U_{\textup{RW}} + U_{\textup{wall}}
\end{equation}
where $U_{\textup{LW}}(x, y) = (x-x_{\textup{LW}})^2 + (y-y_{\textup{LW}})^2$ and $U_{\textup{RW}}(x, y) = (x-x_\text{RW})^2 + (y-y_\text{RW})^2$ are the potentials defining the left and right wells, respectively. The wells' minima are located at $(x_\text{LW}, y_\text{LW})$ and $(x_\text{RW}, y_\text{RW})$, respectively. $U_{\textup{wall}}(r_\text{w}, \phi_\text{w}) = - k_{\text{wh}} \times U_r(r_\text{w}) \times U_{\phi}(\phi_\text{w})$ is the potential describing the wall with $U_{r}(r_\text{w}) = \frac{1}{1+\exp{\big(-k_\text{ws}(r_\text{w}-r_\text{a})\big)}} + \frac{1}{1+\exp{\big(k_\text{ws}(r_\text{w}-r_\text{b})\big)}} -1$ and 
$U_{\phi}(\phi_\text{w}) = \frac{1}{1+\exp{\big(-k_\text{ws}(\phi_\text{w}-\pi k_\text{wc})\big)}} + \frac{1}{1+\exp{\big(k_\text{ws}(\phi_\text{w}-\pi k_\text{wc})\big)}} -1$. This potential is defined as an arc centred around $(x_\text{LW}, y_\text{LW})$ and is parameterised by $(r_\text{w}, \phi_\text{w})=(\sqrt{(x-x_{\text{LW}})^2 + (y-y_{\text{LW}})^2},\, \text{atan2}(y-y_{\text{LW}}, x-x_{\text{LW}}))$. The wall's height is dictated by $k_\text{wh}$ and its steepness by $k_\text{ws}$. 
The wall's distance from the left well, $d$, and its width, $w$, define $r_\text{a}=d-w/2$ and $r_\text{b}=d+w/2$.
What proportion of a full circle the arc of the wall covers is controlled by $k_\text{wc}$.
We set $(x_\text{LW}, y_\text{LW})=(-1,0)$, $(x_\text{RW}, y_\text{RW})=(1,0)$, $k_\text{wh}=100$, $k_\text{ws}=10$, $d=1.5$, $w=0.3$ and $k_\text{wc}=0.1$.
Simulations on this landscape all begin from the minimum of the left well and the target region encompassed all points where $x>0$ and $U(x,y) \leq 1$.
The start point and target region are shown in Fig. \ref{fig:Fig 3}c.
Particles use the same dynamics as the active particles on the double-well described above, but with $\Delta t=0.025$ (finer time resolution required to avoid numerical instability near the steep wall), $v=2$, and variable $D_\theta$ to test different Pe\'clet numbers (i.e. different levels of activity). 

\subsubsection*{Dual-Channel Double-Well Potential}
This potential (Fig. \ref{fig:Fig 4}a) consists of two wells connected by two distinct channels: a top channel and a bottom channel.
A peak of variable height separates the two. 
Along the top channel lies a barrier of height 1.1 $k_\text{B}T_\text{eff}$, and along the bottom channel a barrier of 0.9 $k_\text{B}T_\text{eff}$.
\begin{equation}\label{eq:TCDW}
    U(x,y) = 9\Big(x^4 - 2y^2 + y^4 +\frac{78}{37}x^2(y^2-1) + \frac{1}{90}y + 1.11097 + k_\text{BH}\ \exp{(-(x^2+y^2))}\Big)
\end{equation}

\noindent The height of the peak between the two channels can be modulated by changing $k_\text{BH}$.
Our simulations used $k_\text{BH}=0$ for a resultant peak height of 10 $k_\text{B}T_\text{eff}$,  $k_\text{BH}=5/9$ for 15 $k_\text{B}T_\text{eff}$, $k_\text{BH}=10/9$ for 20 $k_\text{B}T_\text{eff}$, $k_\text{BH}=5/3$ for 25 $k_\text{B}T_\text{eff}$, and $k_\text{BH}=20/9$ for 30 $k_\text{B}T_\text{eff}$. Simulations on this potential all began from the minimum of the left well, $(-1.\overline{03},0)$.
The target region encompasses all points where $x>0$ and $U(x,y) \leq 0.25$.
The start point and target region are shown in Fig. \ref{fig:Fig 4}a.
Particles move according to Eqs. \ref{eq:ABP_1} and \ref{eq:ABP_2}; we set $\Delta t=0.05$, $v=0.15$, $\mu=0.25$, $D=0.1$, $D_{\theta}=1$ and $i_\textup{max}=32$.

\subsection*{FlowRES Network Architecture}
The neural network FlowRES uses is a normalising flow (see Normalizing Flows),\cite{NFrevGP,NFReviewIK} implemented with affine coupling transformations\cite{RealNVP} (see Affine Coupling Layers)  that employ WaveNet (see section on WaveNet below)\cite{WaveNet} as part of their architecture.
We describe our overall flow architecture in terms of \textit{scales}  (see Multi-scale Architecture) and \textit{complete flow steps} (Figs. \ref{fig:SupShematic_PosOnly}a and \ref{fig:SupShematic_Full}a). As shown in Fig. \ref{fig:SupShematic_PosOnly}a, networks for passive systems are composed of two scales, each with ten passive complete flow steps. 
The network architecture for active systems is shown in Fig. \ref{fig:SupShematic_Full}a, where we have two scales with five active complete flow steps. 
At each scale, an input position matrix is reshaped (Fig. \ref{fig:SupShematic_PosOnly}b) before being split into two channels (Fig. \ref{fig:SupShematic_PosOnly}c) which are passed to several consecutive complete flow steps.
After the complete flow steps, one of the two channels is factored out, while the other moves forward to the next scale.
All channels that have been factored out at each scale are recombined to obtain the final output.
In addition to the aforementioned position matrix, for active systems we have an orientation matrix (Fig. \ref{fig:SupShematic_Full}); at each scale, this is also reshaped and split into two channels, with one channel being factored out. 
For passive systems, a complete flow step is composed of an affine coupling block (i.e., a pair of alternating affine coupling layers Fig. \ref{fig:SupShematic_PosOnly}d-e) sandwiched between a $1\times1$ convolution and its inverse (see $1x1$ Convolutions). 
For active systems, the paths we generate must be conditioned by orientation walks. To do so, active complete flow steps are first composed of two affine coupling blocks, each sandwiched between a $1\times1$ convolution and its inverse; after this, the two position channels are recombined via an inverse split operation with the combined position channel fed alongside the orientation walk into an angle cross-coupling layer (see Angle Cross-Coupling) (Fig. \ref{fig:SupShematic_Full}b-c).

\subsubsection*{Normalizing Flows}
Normalising flows push a simple unimodal base distribution through a series of transformations to produce a richer multimodal distribution.\cite{NFrevGP} Our normalizing flow network $\textbf{\textit{F}}$ transforms the $i_\text{max}\times n$ ($n=2$ for passive particles, $n=3$ for active particles) base matrices $\textbf{w}_\text{B}$ to generate the distribution over the transition paths $\textbf{w}'$ (Figs. \ref{fig:SupShematic_PosOnly} and \ref{fig:SupShematic_Full}): 

\begin{equation}
    \textbf{w}' = \textbf{\textit{F}}(\textbf{w}_\text{B}) \quad\text{where}\quad \textbf{w}_\text{B} \sim \rho_{\text{B}}(\textbf{w}_\text{B})
\end{equation}

\noindent with $\rho_{\text{B}}(\textbf{w}_\text{B})$ being the simple unimodal base distribution that the base matrices $\textbf{w}_\text{B}$ are sampled from. For passive particles, $\textbf{w}_\text{B}$ are $i_\text{max} \times 2$-matrices of random values sampled from an uncorrelated multivariate Gaussian distribution. For active particles, $\textbf{w}_\text{B}$ are $i_\text{max} \times 3$-matrices created by horizontally concatenating $i_\text{max} \times 2$ random Gaussian matrices with $i_\text{max} \times 1$ sequences of orientations following Eq. \ref{eq:ABP_2}. 

The defining feature of flow-based models is that $\textbf{\textit{F}}$ must be invertible ($\textbf{w}_\text{B} = \textbf{\textit{F}}^{-1}(\textbf{w}))$ and both $\textbf{\textit{F}}$ and $\textbf{\textit{F}}^{-1}$ must be differentiable.
Such functions are known as diffeomorphisms.\cite{diffeomorphisms}
Under these conditions, the probability of $\textbf{w}'$ is well-defined and can be obtained by a change of variables: \cite{changeOFvariables}
\begin{equation}\label{eq: density}
    \rho_{\text{NF}} (\textbf{w}') = \rho_\text{B}(\textbf{w}_\textup{B}) |\det J_{\textbf{\textit{F}}}(\textbf{w}_\text{B})|^{-1} 
                      = \rho_\text{B}(\textbf{\textit{F}}^{-1}(\textbf{w}')) |\det J_{\textbf{\textit{F}}^{-1}}(\textbf{w}')|
\end{equation}

\noindent
where $J_{\textbf{\textit{F}}}(\textbf{w}_\text{B})$ and $J_{\textbf{\textit{F}}^{-1}}(\textbf{w}')$ are the Jacobian matrices of $\textbf{\textit{F}}$ and $\textbf{\textit{F}}^{-1}$, respectively.
Efficient calculation of $\rho_{\text{NF}}(\textbf{w}')$ necessitates efficient calculation of $J_{\textbf{\textit{F}}}(\textbf{w}_\text{B})$ and $J_{\textbf{\textit{F}}^{-1}}(\textbf{w}')$, which is made possible by constructing $\textbf{\textit{F}}$ using affine coupling layers (next section).

Flow-based models are trained, like other probabilistic models, through minimisation of the divergence between the distribution they generate and the distribution we are training them to generate.
This minimisation occurs with respect to the parameters of $\textbf{\textit{F}}$, $\psi$, which are the weights of our normalising flow - specifically of the $s$ and $t$ networks, and $1\times1$ convolutions.
Training proceeds using the Kullback-Leibler divergence,\cite{NFrevGP} most commonly implemented with the maximum likelihood principle. \cite{BoltzmannGenerators}
Paths $\textbf{w}$ from the FlowRES distribution $\hat\rho(\textbf{\text{w}})$ are transformed into their base space representations $\textbf{\textit{F}}^{-1}(\textbf{w})=\textbf{w}_\text{B}$.
Maximizing the likelihood of $\textbf{w}_\text{B}$ in the base distribution $\rho_\text{B}$ corresponds to minimizing the loss function:
\begin{equation}\label{L_ML}
    \mathcal{L}(\psi) = \mathbb{E}_{\textbf{w}}[\log(\rho_\text{B}(\textbf{\textit{F}}^{-1}(\textbf{w}))) - \log|\det J_{\textbf{\textit{F}}^{-1}}(\textbf{w})|]
\end{equation}

\noindent The FlowRES distribution $\hat\rho(\textbf{\text{w}})$ converges to the target distribution $\rho_{\rm T}(\textbf{\text{w}})$ with each iteration $m$.
Therefore, as $m$ grows, $\mathcal{L}(\psi)$ minimises the divergence between the network proposal distribution $\rho_\text{NF}(\textbf{w}')$ and the target $\rho_{\rm T}(\textbf{\text{w}})$.
Note that our loss function uses the reverse flow $\textbf{\textit{F}}^{-1}$, transforming walks $\textbf{\textup{w}}$ into basis matrices $\textbf{\textup{w}}_B$, but generating new paths uses the forward flow $\textbf{\textit{F}}$ to transform $\textbf{\textup{w}}_B$ into new proposals $\textbf{\textup{w}}'$.

\subsubsection*{Affine Coupling Layers}
Affine coupling layers are powerful, yet computationally efficient reversible transformations with easy to calculate log-determinants (Eq. \ref{L_ML}), thus are well suited for constructing normalising flows.
They split an input vector into two channels and update one channel with a bijection parameterised by some function of the other (Fig. \ref{fig:SupShematic_PosOnly}d-e). 
The channels are split using an alternating mask wherein all configurations with even indices form one channel and all odd configurations form another (Fig. \ref{fig:SupShematic_PosOnly}c).
Consider an $i\times n$ dimensional matrix $\omega_0$; performing an alternating split on this yields two new matrices: an $\lfloor i/2\rfloor \times n$ matrix $\alpha_0$ and an $i - \lfloor i/2\rfloor \times n$ matrix $\beta_0$:
\begin{equation}
    \text{split}(\omega_0) = \alpha_0, \,\beta_0
\end{equation}

\noindent Passing these matrices to an affine coupling layer will produce outputs $\alpha_1$ and $\beta_1$:
    \begin{equation}\label{NVP_fw}
    \begin{aligned}
    \alpha_1
        & = \alpha_0 \\
    \beta_1
        & = \beta_0\odot \exp{[s(\alpha_0)]} + t(\alpha_0) \\
    \end{aligned}
\end{equation}
The above set of operations is a single affine coupling layer,\cite{RealNVP} wherein $s$ and $t$ stand for scale and translate and are functions from $\mathbb{R}^{\lfloor i/2\rfloor \times n} \mapsto \mathbb{R}^{i - \lfloor i/2\rfloor \times n}$.
The symbol $\odot$ represents Haddamard (element-wise) multiplication. 
Computing the inverse of the coupling layer (used by $\textbf{\textit{F}}^{-1}$) does not require inverting $s$ or $t$,\cite{RealNVP} and is simply given by:
\begin{equation}
    \begin{aligned}\label{NVP_bw}
    \alpha_0
        & = \alpha_1 \\
    \beta_0
        & = (\beta_1 - t(\alpha_1)) \odot \exp{[-s(\alpha_1)]}  \\        
    \end{aligned}
\end{equation}
Combining $\alpha_1$ and $\beta_1$ by inverting the alternating split yields another $i\times n$ matrix, $\omega_1$. The affine coupling transformation has a lower triangular Jacobian matrix, given by:
\begin{equation}
J(\omega_0) = \frac{\partial \omega_1}{\partial \omega_0^{\top}} = 
\begin{bmatrix}
I & 0  \\
\frac{\partial \beta_1}{\partial {\alpha_0}^\top} & \text{diag}(\exp{[s(\alpha_0)}]  \\
\end{bmatrix}\
\end{equation}
\noindent where $\omega_0^{\top}$ and ${\alpha_0}^\top$ denote the transposes of $\omega_0$ and ${\alpha_0}$.
Above,  $\exp{[s(\alpha_0)]}$ represents a matrix whose diagonal elements correspond to each value of $\exp{[s(\alpha_0)]}$. 
Therefore, the Jacobian determinant of a coupling layer can be efficiently calculated as $\exp{\big[\sum s(\alpha_0)}\big]$. 
Note that we never need to invert or calculate the Jacobian determinants of $s$ or $t$, so these functions can be arbitrarily complex; any neural network architecture (here, WaveNet\cite{WaveNet}) may be used for $s$ and $t$.\cite{RealNVP} 
For the sake of training stability, we initialise the last layer of each $s$ and $t$ network with zeros so that each affine coupling layer (and thereby the network as a whole) initially performs an identity operation.
The transformations shown so far leave one of the two channels completely unchanged. This is overcome by composing alternating coupling layers with channels swapped, such that both channels are transformed. 
A pair of alternating coupling layers is an \textit{affine coupling block}. Fig. \ref{fig:SupShematic_PosOnly}d shows a schematic representation of an affine coupling block (used by $\textbf{\textit{F}}$), while Fig. \ref{fig:SupShematic_PosOnly}e shows an inverse affine coupling block (used by $\textbf{\textit{F}}^{-1}$).

\subsubsection*{WaveNet}
As we aim to generate paths, we select an architecture for our $s$ and $t$ networks that is optimised for handling time series data: WaveNet.\cite{WaveNet}
Originally designed for audio prediction, this architecture uses dilated convolutions, gated activation units, and skip connections to extract information from sequential data.\cite{WaveNet, WaveNetLowRateSpeech, WaveNetUnsupervisedSpeech}
The use of a feed forward, convolutional approach like WaveNet, instead of a recurrent neural network based approach precludes sequential dependencies between outputs enabling faster, more stable training and inference.\cite{DifficultyTrainingRNNs}
Each WaveNet uses convolutions with a kernel size of 3, with 3 dilated convolutional layers. 
Networks for passive systems have 32 filters, while 64 filters were required to efficiently address the added complexities of active systems.
For the sake of training stability, we initialise the last layer of each $s$ and $t$ network with zeros so that each affine coupling layer initially performs an identity operation.
All other weights used Glorot normal and He normal initialisation.\cite{Glorot,He}
We deviate from the original WaveNet architecture by using non-causal convolutions as opposed to causal convolutions as we always have access to each whole trajectory (i.e. both past and future time-steps are available for convolution).

\subsubsection*{Multi-scale Architecture}
Propagating large matrices through the whole network would incur a high computational and memory cost. 
To remedy this we factor out channels at each scale (Figs. \ref{fig:SupShematic_PosOnly}a and \ref{fig:SupShematic_Full}a), allowing us to create larger and more efficient models. 
Forcing the network to Gaussianize the channels at different scales also allows for the definition of different levels of representation: channels factored out later contribute coarser features of each generated path, while those factored out earlier contribute more local, fine-grained details.\cite{RealNVP}
Using multiple scales also has the benefit of distributing the loss function throughout the network, stabilizing training.\cite{RealNVP}

\subsubsection*{$\textbf{1} \times \textbf{1}$ Convolutions}
The success of an affine coupling layer depends in part on how its input matrices are partitioned into channels.
Each scale includes an alternating split (Fig. \ref{fig:SupShematic_PosOnly}c) but we then feed the two resultant channels into an invertible $1\times1$ convolution before each affine coupling block (Fig. \ref{fig:SupShematic_PosOnly}a).\cite{GLOW}
This allows for trainable, intelligent mixing between the two channels, leading to more effective transformations by the subsequent coupling block. 
The weight matrix of this convolution is initialised as a random rotation matrix.
After the coupling block, we invert this $1\times1$ convolution, as we found this to improve network performance. 
$1\times1$ convolutions disrupt the time ordering of path matrices, potentially diminishing the efficacy of downstream coupling blocks' processing; we attribute the improvement in performance from inverting the convolutions to the fact that doing so allows for the time ordering of path matrices to be better preserved between coupling blocks. 

\subsubsection*{Angle Cross-Coupling}
As orientation $\theta$ is independent of position $\textbf{r}$ in the active dynamics we consider (Eqs. \ref{eq:ABP_1} and \ref{eq:ABP_2}), the orientation component of any path can be an independently generated random walk. The flow $\textbf{\textit{F}}$ transforms the Gaussian components of each base matrix $\textbf{w}_\text{B}$ into the positional information of the generated paths $\textbf{w}'$; the orientation components condition the aforementioned transformation, but are themselves unchanged.
Our flow begins by splitting the positional component of the base matrix into two channels, and feeding these into a certain number of affine coupling blocks where the two positional channels affect one another (Fig. \ref{fig:SupShematic_Full}a).
Then, we recombine the two position channels into a single channel. 
A single affine coupling layer, parameterised by the orientation data, operates on this combined position channel and ensures that the relationship between the orientation and position data is physically realistic.
We call this an \textit{affine cross-coupling layer} (Fig. \ref{fig:SupShematic_Full}b-c).
Instead of passing raw orientation angles $\theta$ to the $s$ and $t$ layers (again implemented with WaveNet) of this cross-coupling layer, we encode them as a vector $[\sin(\theta),\ \cos(\theta)]$.
This encoding better captures the cyclical nature of angles, being free of any discontinuities across the range of all possible orientations. 

\subsection*{Generation of Transition Ensembles via Direct Integration}
Direct integration of the Langevin dynamics equations (Eqs. \ref{eq:ABP_1} and \ref{eq:ABP_2}) is used to benchmark FlowRES and to generate validation ensembles.
In both cases, paths are generated for a given system up to $i_\text{max}$ time-steps.
If any of a path's microstates $w_i$ enters the system's target region we consider it a transition path. 
Ensembles used for benchmarking are composed of 10,000 transition paths, while validation ensembles consist of 50,000 transition paths. 

\subsection*{Jensen Shannon Divergence}

For validation purposes only, we measure the quality of any generated ensemble by calculating its similarity to a validation ensemble composed of 50,000 paths from the target distribution generated via direct integration of the Langevin equations (Eqs. \ref{eq:ABP_1} and \ref{eq:ABP_2}).
The similarity between these ensembles can be quantified by calculating the Jensen-Shannon distance (JSD) between the probability densities $\hat{\rho}(x,y)$ (the generated ensemble) and ${\rho}_{\rm V}(x,y)$ (the validation ensemble):
\begin{equation}
    \mathrm{JSD}(\hat{\rho}(x,y)||\rho_{\rm V}(x,y)) = \sqrt{\frac{{D_{\rm KL}}\big(\hat{\rho}(x,y)||M\big)+{D_{\rm KL}}\big(\rho_{\rm V}(x,y)||M\big)}{2}}
\end{equation}
\noindent where $M=1/2\big(\hat{\rho}(x,y)+\rho_{\rm V}(x,y)\big)$ and $D_{\rm KL}$ is the Kullback-Leibler divergence.\cite{KLD} 
As a baseline for comparison, we first generate 10,000 paths by direct integration, for which the computational cost is measured in terms of required number of path proposals and CPU time (Figs. \ref{fig:Fig 2}e-g, \ref{fig:Fig 3}f-g, \ref{fig:SupABPActive} and \ref{fig:SupABPTCDW}b-c). 
We then calculate the JSD between these 10,000 paths and the validation ensemble, and establish this as a target JSD.
Finally, to evaluate the efficiency of FlowRES, we monitor the JSD between each iteration of FlowRES distribution and our validation distribution.
When this JSD is lower than the target JSD, we can conclude that the FlowRES-generated ensemble is at least as statistically useful as 10,000 direct integration paths.
This allows us to compare the computational cost of generating paths with each method.

\subsection*{Computational Resources}
All computation used the same node on the Nanyang Technology University's Gekko cluster with an Intel Skylake Xeon Gold 6150 processor and Nvidia V100 GPU for network training.

\section*{Acknowledgments}
The authors thank Francesco Gervasio for critical reading of the manuscript. SA, QXP and GV are grateful to the studentship funded by the A*STAR-UCL Research Attachment Programme through the EPSRC M3S CDT (EP/L015862/1). RN acknowledge the support by the Singapore Ministry of Education through the Academic Research Tier 2 Fund (MOE2019-T2-2-010) and Tier 1 grant (RG59/21).

\section*{Author contributions}
Author contributions are defined based on the CRediT (Contributor Roles Taxonomy) and listed alphabetically. Conceptualization: GV, RN. Data Curation: SA, GV, RN. Formal analysis: SA, GV, RN. Funding acquisition: QXP, GV, RN. Investigation: SA, GV, RN. Methodology: SA, GV, RN. Project administration: QXP, GV, RN. Resources: QXP, GV, RN. Software: SA. Supervision: QXP, GV, RN. Validation: SA. Visualization: SA. Writing – original draft: SA. Writing – review and editing: All.

\section*{Competing interests}

The authors have no competing interests.

\bibliography{references}

\renewcommand{\thefigure}{S\arabic{figure}}
\setcounter{figure}{0}

\newpage
\section*{Supplementary Material}

\begin{figure*}[ht]
\centering
\includegraphics[width=0.775\linewidth]{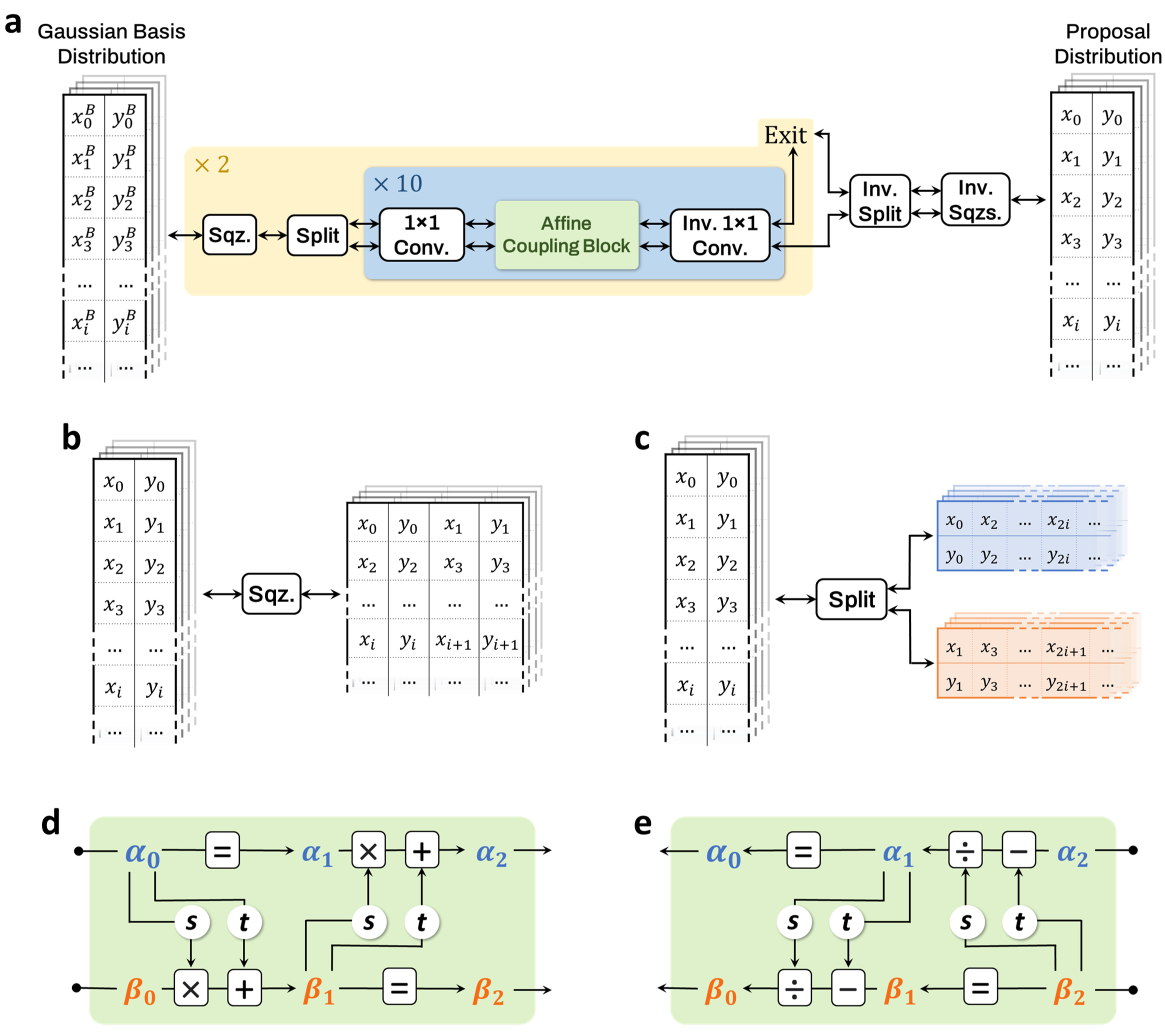}
\caption{\textbf{FlowRES normalising flow network architecture for passive systems.} 
(\textbf{a}) FlowRES network for passive systems. Input Gaussian basis matrices are passed into the first of $2$ \textit{scales} (Methods), which begin by reshaping the input with a squeeze operation (see \textbf{b}) followed by dividing it into two channels with an alternating split (see \textbf{c}). The two channels are passed to $10$ consecutive \textit{complete flow steps}, each composed of an \textit{affine coupling block} (see \textbf{d} and \textbf{e}) preceded by a $1\times1$ convolution and followed by the inverse of this convolution. After the complete flow steps, one of the two channels is factored out while the other channel passes on to the next scale. After both scales, all the squeeze operations and splits are inverted, yielding a distribution of proposed paths.
({\textbf b}) In a squeeze operation, the matrices in a channel all change shape, doubling in width but halving in height. A squeeze is equivalent to an alternating split (see \textbf{c}) followed by the two resulting channels being horizontally concatenated.
({\textbf c}) In an alternating split, a single channel is split into two. Even indices form one channel (blue) and odd indices form another (red).
(\textbf {d}) An affine coupling block used to generate path proposals. The channel $\alpha_0$ is passed into two WaveNet networks (Methods), the outputs of which parameterize a bijective transformation $\beta_1=\beta_0\odot \exp{[s(\alpha_0)]} + t(\alpha_0)$ of the second channel $\beta_0$. This new $\beta_1$ is then passed into other WaveNets, whose outputs are used to parameterize $\alpha_2=\alpha_1\odot \exp{[s(\beta_1)]} + t(\beta_1)$.
(\textbf {e}) An inverted affine coupling block (used for training). 
}
\label{fig:SupShematic_PosOnly}
\end{figure*}

\begin{figure*}[ht]
\centering
\includegraphics[width=\linewidth]{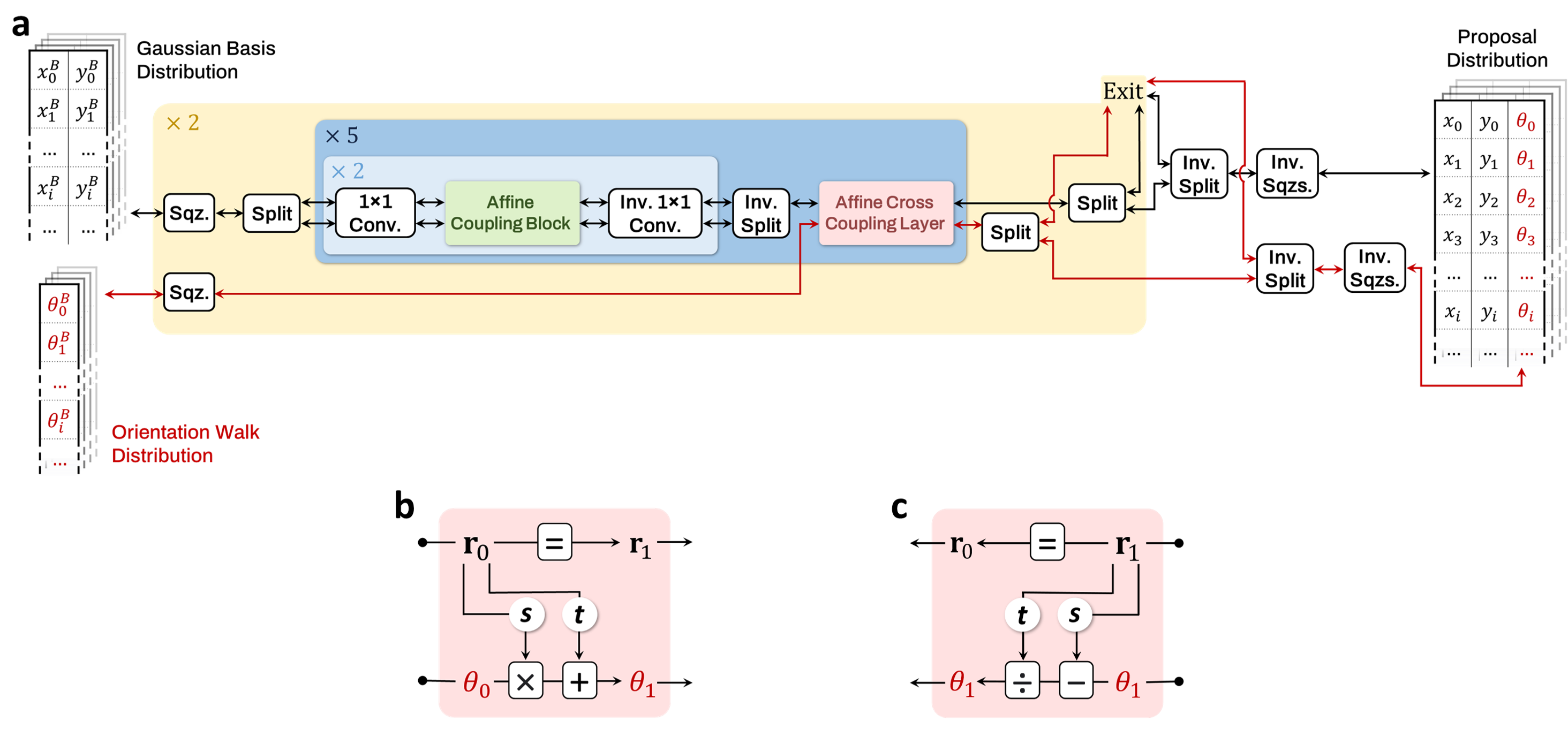}
\caption{\textbf{FlowRES normalising flow network architecture for active systems.}
({\textbf a}) As with the passive case in Fig. \ref{fig:SupShematic_PosOnly}a, input Gaussian basis matrices are passed into the first of $2$ \textit{scales} (Methods); these matrices will be transformed to form the position component of the final proposed paths. Alongside these, the orientations of a random walk are also passed to the scales. Both of these inputs are reshaped with squeeze operations (Fig. \ref{fig:SupShematic_PosOnly}b). The squeezed Gaussian basis matrices are divided into two channels with an alternating split (Fig. \ref{fig:SupShematic_PosOnly}c), and these two position channels (black lines) are then passed to $5$ \textit{complete flow steps}: each step is first composed of two affine coupling blocks sandwiched between $1\times1$ convolutions and inverse $1\times1$ convolutions (Fig. \ref{fig:SupShematic_PosOnly}d-e); after these two affine coupling blocks, the two position channels are recombined and fed alongside the orientation walk into an angle cross-coupling layer (see \textbf{b-c}).
A position channel and orientation channel are the output of this coupling layer.
Each of these then undergo alternating splits (Fig. \ref{fig:SupShematic_PosOnly}c), with one channel from each being factored out and the other moving onward to the next scale.
After both scales, all the squeeze operations and splits are inverted. Combining the position and orientation channels yields a distribution of proposed paths. 
({\textbf b}) An affine cross-coupling layer used to generate path proposals. The orientation channel $\theta$ is passed into two WaveNet networks (Methods), the outputs of which parameterize a bijective transformation that updates the position channel: $\textbf{r}_1=\textbf{r}_0\odot \exp{[s(\theta_0)]} + t(\theta_0)$. Note that the orientation channel itself remains unchanged by this layer. (\textbf {c}) An inverted affine cross-coupling layer, used for training.}
\label{fig:SupShematic_Full}
\end{figure*}

\begin{figure*}[ht]
\centering
\includegraphics[width=0.8\linewidth]{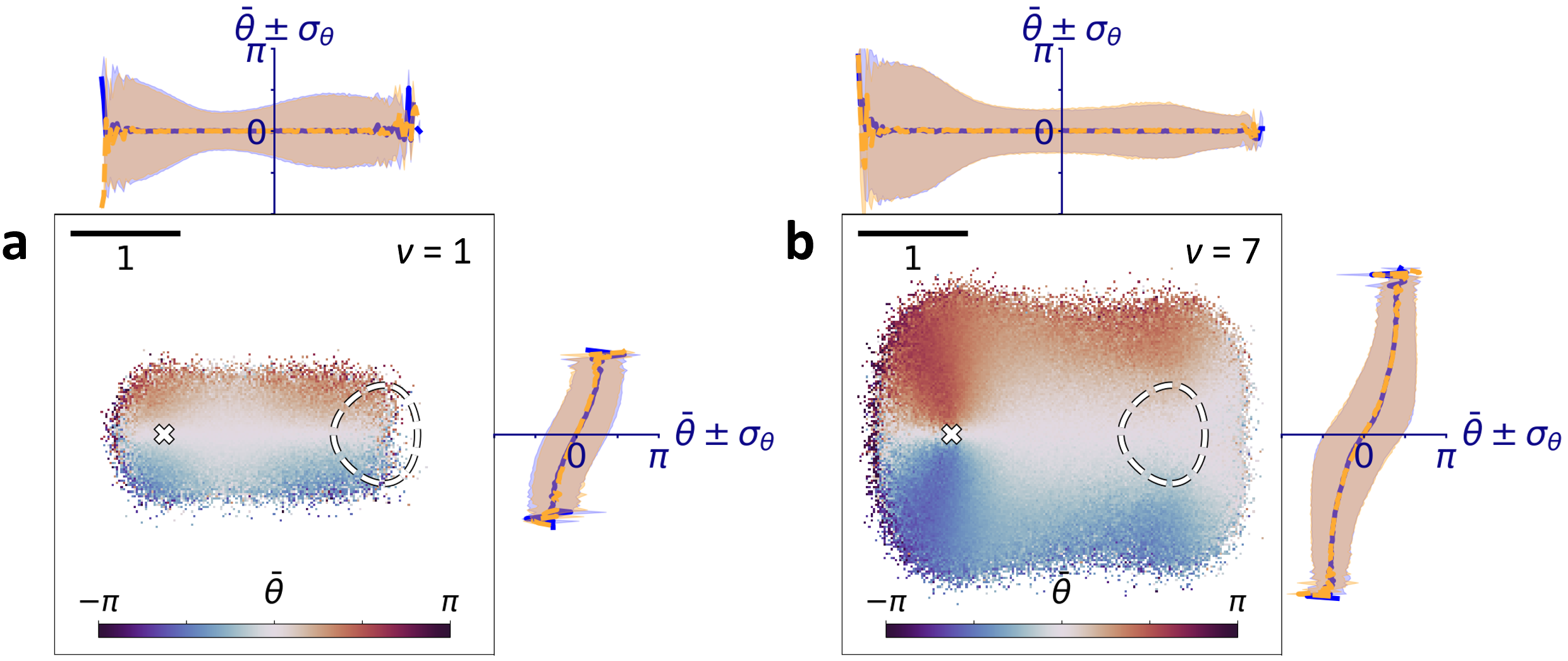}
\caption{\textbf{Orientations of active Brownian particles for barrier crossing in a double-well potential.}
The color maps show FlowRES-generated average orientations $\overline{\theta}$ at different points on a double-well potential (Fig. \ref{fig:Fig 2}a, barrier height of $15 \, k_{\rm B} T_{\rm eff}$ as in Fig. \ref{fig:Fig 3}a-b) for target-reaching active Brownian particles with (\textbf{a}) $v=1$ and (\textbf{b}) $v=7$. These are flanked by lines showing the average orientations at different $x$ (above) and $y$ (right) values (standard deviations given as shaded regions) generated by FlowRES (blue lines and shaded blue regions) and by direct integration (dashed yellow lines and shaded yellow regions) of Langevin equations (Eqs. \ref{eq:ABP_1} and \ref{eq:ABP_2}). Note the strong correspondence between both results. Starting point (white cross) and target region (dashed line) are shown for reference. Unitary scale bar: upper left. Simulation parameters as in Fig. \ref{fig:Fig 3}a-b.}
\label{fig:SupABPActive}
\end{figure*}

\begin{figure*}[ht]
\centering
\includegraphics[width=0.8\linewidth]{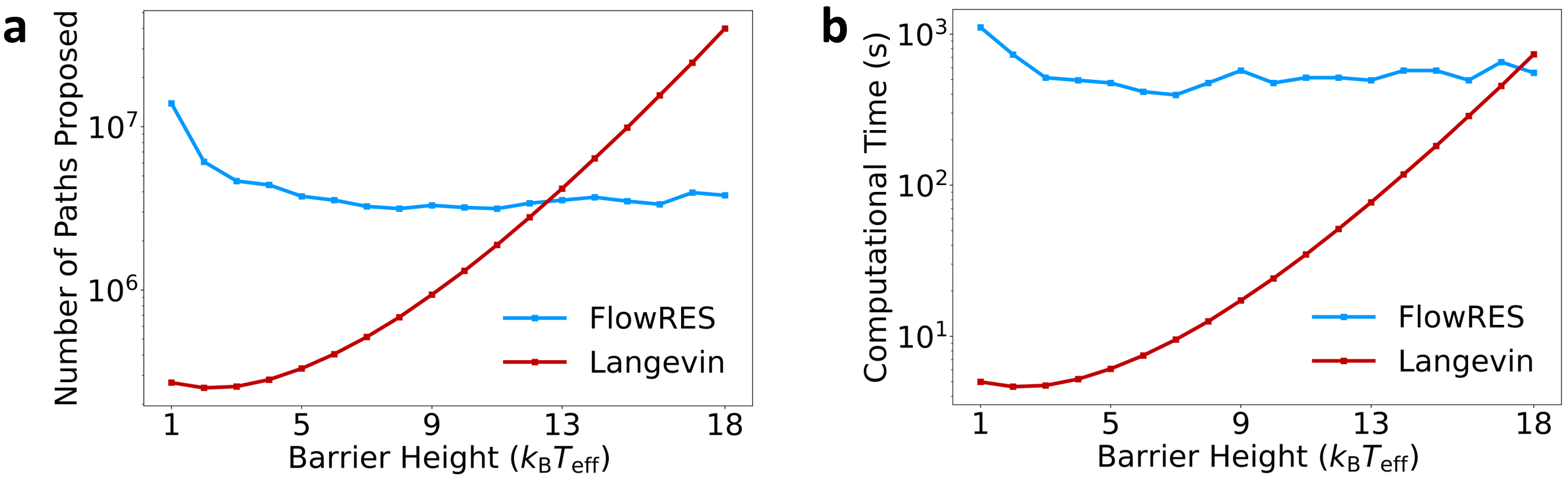}
\caption{\textbf{Computational efficiency for barrier crossing of active Brownian particles in a double-well potential.}
({\textbf a}) Total number of active particles' paths ($v=1$) proposed to yield 10,000 target-reaching paths with null JSD deviation (Methods) and ({\textbf b}) corresponding computational time with FlowRES (blue line) and direct integration (red line) as a function of barrier height of the double-well potential in Fig. \ref{fig:Fig 2}a.
The computational cost of direct integration of Langevin equations increases exponentially with barrier height, while FlowRES maintains near constant computational cost. Simulation parameters as in Fig. \ref{fig:Fig 3}a.}
\label{fig:SupABPActive_scale}
\end{figure*}

\begin{figure*}[ht]
\centering
\includegraphics[width=0.8\linewidth]{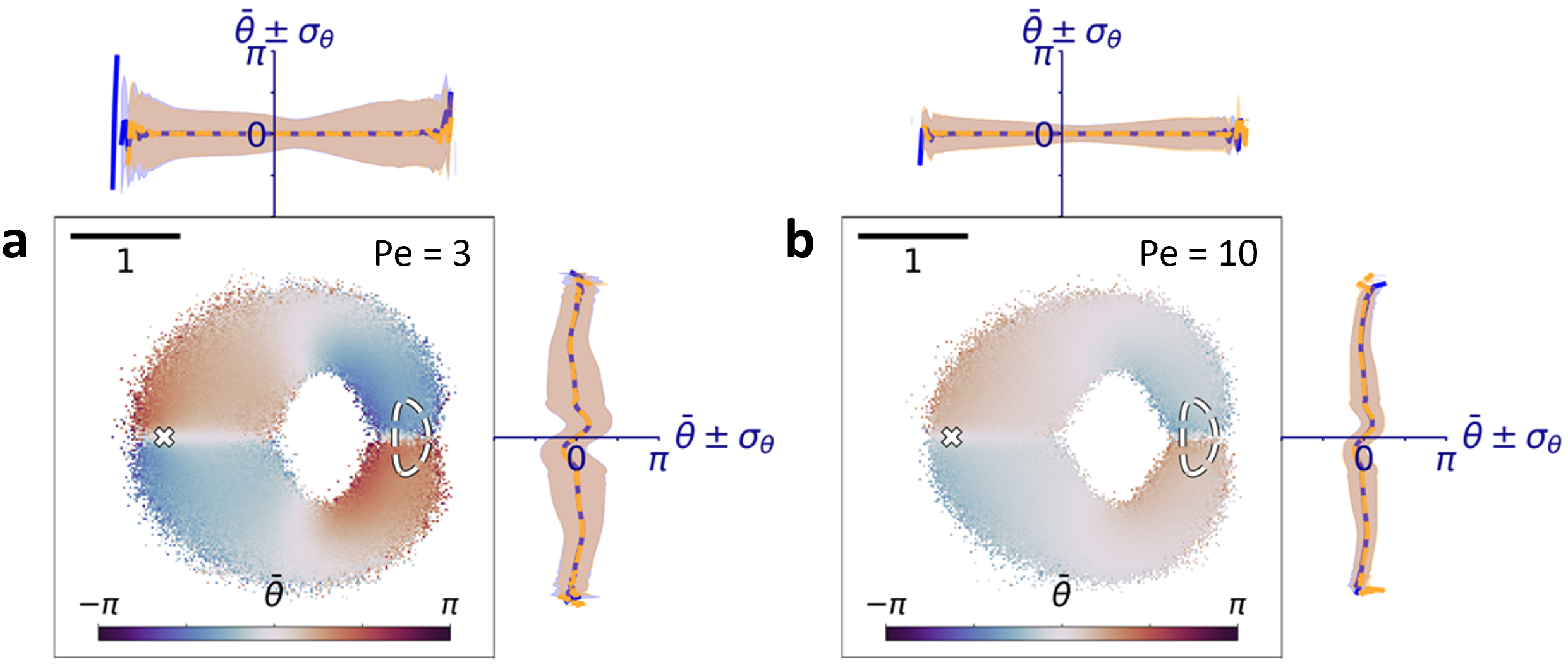}
\caption{\textbf{Orientations of active Brownian particles for circumventing an insurmountable wall on a double-well potential.}
The color maps show FlowRES-generated values for average orientations $\overline{\theta}$ at different points on a double-well potential with an insurmountable wall separating them (Fig \ref{fig:Fig 3}c, Methods) for target-reaching active Brownian particles with (\textbf{a}) ${\rm Pe} = 3$ and (\textbf{b}) ${\rm Pe} = 10$. These are flanked by lines showing the average orientations at different $x$ (above) and $y$ (right) values (standard deviations given as shaded regions) generated by FlowRES (blue lines and shaded blue regions) and by direct integration (dashed yellow lines and shaded yellow regions) of Langevin equations (Eqs. \ref{eq:ABP_1} and \ref{eq:ABP_2}).  Note the strong correspondence between both results. Starting point (white cross) and target region (dashed line) are shown for reference. Unitary scale bar: upper left. Simulation parameters as in Fig. \ref{fig:Fig 3}d-e.
}
\label{fig:SupABPBlock}
\end{figure*}

\begin{figure*}[ht]
\centering
\includegraphics[width=\linewidth]{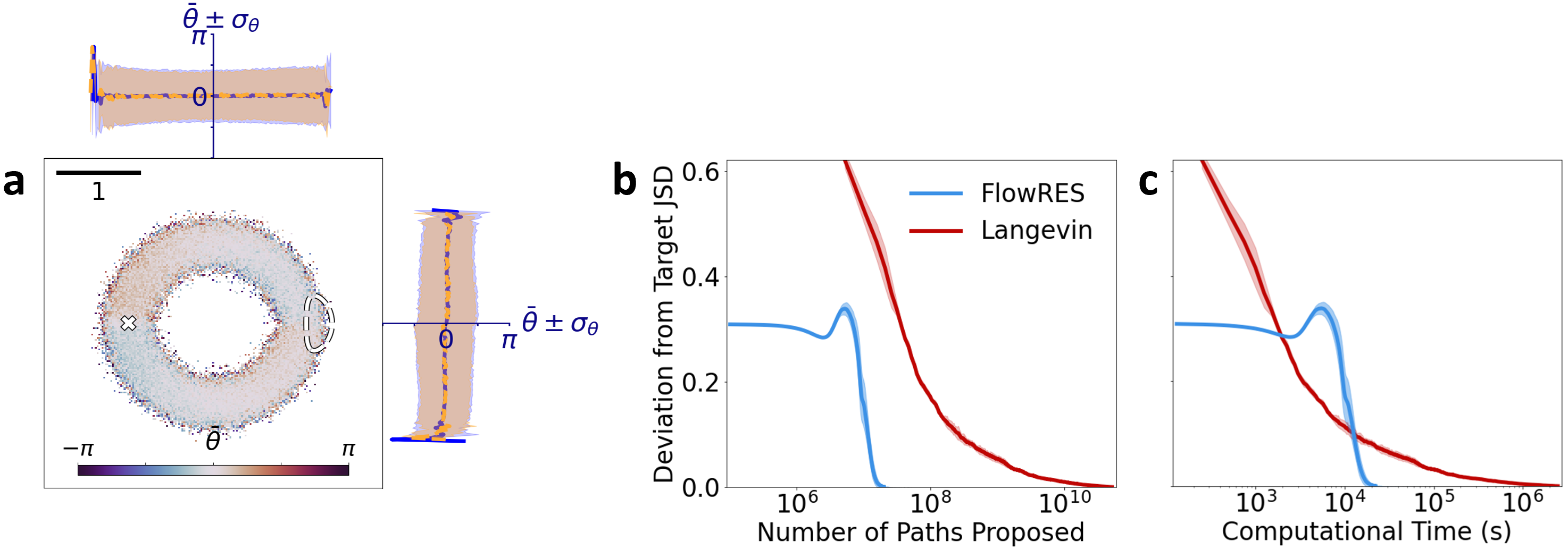}
\caption{\textbf{Enhanced sampling of active Brownian particles exploring an asymmetric dual-channel double-well potential.}
(\textbf{a}) The color map shows FlowRES-generated values for the average orientations $\overline{\theta}$ at different points on a dual-channel double-well potential (Fig. \ref{fig:Fig 4}a, Methods) for target-reaching active Brownian particles. This is flanked by lines showing the average orientations at different $x$ (above) and $y$ (right) values (standard deviations given as shaded regions) generated by FlowRES (blue lines and shaded blue regions) and by direct integration (dashed yellow lines and shaded yellow regions) of Langevin equations (Eqs. \ref{eq:ABP_1} and \ref{eq:ABP_2}). Note the strong correspondence between both results. Starting point (white cross) and target region (dashed line) are shown for reference. Unitary scale bar: upper left. Simulation parameters as in Fig. \ref{fig:Fig 4}.
(\textbf{b-c}) Average deviations of the Jensen Shannon distance (JSD) between validation ensemble and
ensembles of target-reaching paths generated with FlowRES (blue lines) and direct integration (red lines) from the target JSD
(Methods) as a function of (\textbf{b}) the number of paths proposed and (\textbf{c}) the computational time.
Shaded areas represent standard deviations around the average from five independent simulations.}
\label{fig:SupABPTCDW}
\end{figure*}

\end{document}